An ecological framework for the analysis of prebiotic chemical reaction networks and their dynamical behavior

**Authors**

Zhen Peng[1], Alex Plum[1,2], Praful Gagrani[1,3], David A. Baum[1,4]*

**Affiliations**

[1]Wisconsin Institute for Discovery, University of Wisconsin-Madison, Madison WI 53706, USA

[2]Department of Engineering Physics, University of Wisconsin-Madison, Madison WI 53706, USA

[3]Department of Physics, University of Wisconsin-Madison, Madison WI 53706, USA

[4]Department of Botany, University of Wisconsin-Madison, Madison WI 53706, USA

*Correspondence to: David A. Baum (dbaum@wisc.edu).

**Abstract**

It is becoming widely accepted that very early in life's origin, even before the emergence of genetic encoding, reaction networks of diverse small chemicals might have manifested key properties of life, namely self-propagation and adaptive evolution. To explore this possibility, we formalize the dynamics of chemical reaction networks within the framework of chemical ecosystem ecology. To capture the idea that life-like chemical systems are maintained out of equilibrium by fluxes of energy-rich food chemicals, we model chemical ecosystems in well-mixed containers that are subject to constant dilution by a solution with a fixed concentration of food chemicals. Modelling all chemical reactions as fully reversible, we show that seeding an autocatalytic cycle with tiny amounts of one or more of its member chemicals results in logistic growth of all member chemicals in the cycle. This finding justifies drawing an instructive analogy between an autocatalytic cycle and the population of a biological species. We extend this finding to show that pairs of autocatalytic cycles can show competitive, predator-prey, or mutualistic associations just like biological species. Furthermore, when there is stochasticity in the environment, particularly in the seeding of autocatalytic cycles, chemical ecosystems can show complex dynamics that can resemble evolution. The evolutionary character is especially clear when the network architecture results in ecological precedence ("survival of the first"), which makes the path of succession historically contingent on the order in which cycles are seeded. For all its simplicity, the framework developed here is helpful for visualizing how autocatalysis in prebiotic chemical reaction networks can yield life-like properties. Furthermore, chemical ecosystem ecology could provide a useful foundation for exploring the emergence of adaptive dynamics and the origins of polymer-based genetic systems.

**INTRODUCTION**

Empirical and theoretical analyses during the past few decades have converged on the view that the origin of life might entail spontaneous, life-like behavior emerging in networks of relatively simple chemical reactions (Hordijk et al., 2010; Kauffman, 1986; Ruiz-Mirazo et al., 2014; Shapiro, 2006; Walker and Davies, 2013). However, despite a growing body of theory, it is still unclear how simple chemical rules gave rise to systems manifesting the basic properties of life, namely self-propagation and the capacity for adaptive evolution (Joyce, 1994; Luisi, 1998). While abstract models have shown that autocatalysis, the chemical equivalent of self-propagation, is likely to be a common feature of chemical networks underlying the origin of life (Hordijk et al., 2012; Hordijk and Steel, 2016, 2004; Virgo et al., 2016; Virgo and Ikegami, 2013; Xavier et al., 2019), these models have, by-and-large, lacked realistic chemical kinetics, making it difficult to connect their theory of autocatalysis to plausible prebiotic settings. Furthermore, there has been relatively little work on how and under what conditions chemical reaction networks can exhibit dynamics indicative of "evolution" (Goldford and Segrè, 2018; Hordijk et al., 2012; Vasas et al., 2010).

Here we adopt the position that ecological theory provides a framework that can be applied to elucidate the dynamical behavior of complex systems composed of multiple interacting autocatalytic subsystems. Given the assumption that the first form of life was a complex network of chemical reactions that was kept away from thermodynamic equilibrium by a sustained flux of a set of food chemicals, the network and its environment can be viewed as an ecosystem, with autocatalytic subsystems of the network functioning as "biological species". In this ecosystem-first perspective (Baum, 2018; Hunding et al., 2006; Wieczorek, 2012), early life is an open ecosystem of interacting actual or potential autocatalytic subsystems that can show long term changes as a result of its internal dynamics, environmental changes, and the rare influx of new network components from other environments. Although, here, we will only analyze the earliest stages of self-organization with small numbers of autocatalytic subsystems arising and interacting, we believe that this basic model of prebiotic chemical ecosystems can be readily expanded to include polymers and specific catalysts.

In this paper, we develop a model of autocatalytic chemical reaction networks that differs from most existing models in the origin-of-life field (Hordijk and Steel, 2016, 2004; Kauffman, 1993, 1986; Steel, 2000) by assuming that all reactions are reversible and follow conventional mass-action chemical kinetics, without a need for specific catalysts. Using this model, we will show that a single autocatalytic system exhibits dynamics similar to the population dynamics of a single biological species. Furthermore, the interactions between multiple autocatalytic systems can be described in the framework of community ecology, including competitive, predator-prey, and mutualistic interactions. Chemical ecosystems can manifest succession, changing in predictable patterns, with a general tendency to more efficiently utilize resources over time. Finally, we show that rare stochastic perturbations can move chemical ecosystems to new quasistable states, meaning that the state of a system can be said to manifest a memory of environments past, implying heritability and the possibility of adaptive evolution.

Our study suggests that systems chemists may benefit from applying principles of ecosystem ecology and, conversely, that many familiar ecological principles can be understood more fundamentally in physicochemical terms. As a result, we believe that the conceptual framework of chemical ecosystem ecology can sharpen and guide future research aimed at understanding how chemical reaction networks on the early Earth could have complexified over time to eventually give rise to cells with sophisticated metabolic and genetic systems.

**RESULTS**

**1. The basic model of an autocatalytic cycle**

Autocatalysis is a chemical reaction (Hordijk, 2017), or a sequence of chemical reactions, where at least one chemical is present in both the reactants and products, but with a smaller stoichiometric coefficient on the reactant side than that on the product side (Figure 1A). Such a chemical will be referred to here as a member of the autocatalytic system (Figure 1A, M). Chemicals only present in the reactants will be referred to as food (Figure 1A, F), and those only present in the products will be referred to as waste (Figure 1A, W). Thus, autocatalytic systems consume food to produce more members (and, perhaps, waste).

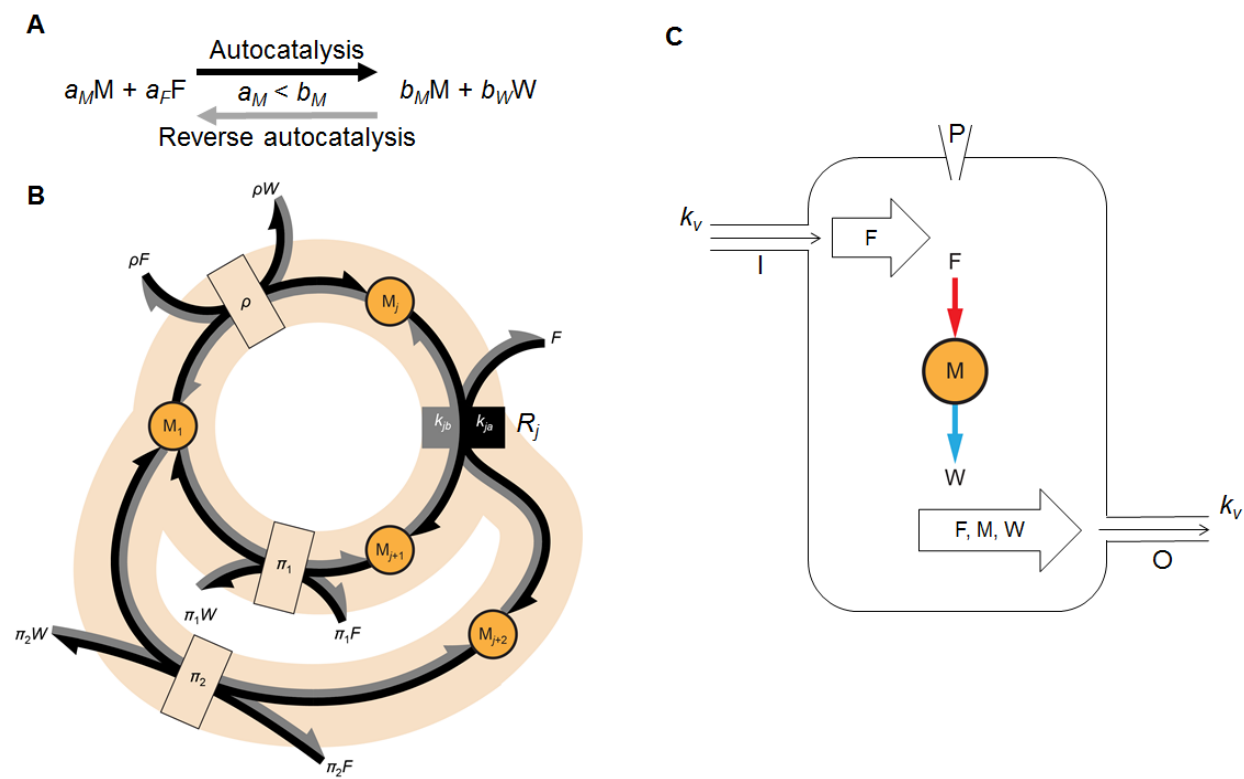

**Figure 1. Autocatalytic cycle and flow reactor**. **(A)** Overall reaction constituting autocatalysis and reverse autocatalysis. *a*'s and *b*'s are stoichiometric coefficients of chemicals on the two sides of a reaction or a sequence of reactions. An autocatalytic process is characterized by at least one chemical (M) being present on both sides and its stoichiometric coefficient on the reactant side being smaller than that of the product side (e.g., $a_M < b_M$). The black arrow indicates autocatalysis; the grey arrow indicates reverse autocatalysis. **(B)** The topological structure of a simple autocatalytic cycle. Black arrows indicate the autocatalytic direction, and grey arrows indicate the reverse autocatalytic direction. Arrows accompanied by black and grey boxes represent a pair of reversible reactions, where the rate constant of a reaction indicated by arrows with a specific color is shown in the box with the same color. Arrows with an empty box containing the label $\rho$, $\pi_1$, or $\pi_2$ represent sequences of $\rho$, $\pi_1$, or $\pi_2$ reactions. $R_j$ is the branching reaction, where, in the autocatalytic direction, one member $M_j$ is on the reactant side and two members $M_{j+1}$ and $M_{j+2}$ are on the product side. $M_1$ is the reunion member, which is generated by both paths that split at $R_j$. Labels $\rho$, $\pi_1$, and $\pi_2$ denote the number of reactions on the three different paths: $\rho$ for the path from $M_1$ to $R_j$, $\pi_1$ for the shorter path from $R_j$ to $M_1$, and $\pi_2$ for the longer path from $R_j$ to $M_1$. **(C)** Flow reactor settings. An input food solution constantly flows into the reactor through entrance I while the solution in the reactor constantly flows out through exit O. The dilution rate is $k_v$. Members or other chemicals can be added through an additional port, P, at different times, as might be needed to seed autocatalytic cycles. The collection of all members of an autocatalytic cycle is represented by a single circle; the red arrow leads from food to members, and the blue arrow leads from members to waste.

Under the assumption, standard in chemistry, that all reactions are reversible, the reverse of an autocatalytic process consumes members and waste to produce more food. To avoid ambiguity, we define the "food" and "waste" based on the direction of autocatalysis, even though the "waste" for autocatalysis could be viewed as the "food" for reverse autocatalysis. We will use the food and waste designations in reference only to a specified autocatalytic cycle. Thus, we allow that the waste of one autocatalytic cycle could serve as the food of another autocatalytic cycle in the same ecosystem. To distinguish food provided at the ecosystem level from food generated within the ecosystem, we will call the former *input food*.

In this paper, we have opted to focus on autocatalytic processes that consist of one or more reversible second-order elementary reactions with a cyclical organization such that one member of the cycle is capable of doubling with each iteration. For example, we will analyze M + F ↔ 2M, but not 2M + F ↔ 3M or M + F ↔ 2M + 2W (the latter type is considered by some prior autocatalytic models; (Field and Noyes, 1974; Prigogine and Lefever, 1968)). The autocatalytic processes we will consider can be represented by a cycle with one branching reaction (Figure 1B, $R_j$) and one reunion member (Figure 1B, $M_1$). Whereas more complicated, multi-branching cycles are possible, we expect that they will show similar overall dynamics and have chosen here to examine only the simple case. These single-branched cycles only rely on second-order elementary reactions, which are the most widely studied (Chang, 2005, pp. 325–328), and resemble some experimentally studied autocatalytic cycles (Boutlerow, 1861; Breslow, 1959; Morowitz et al., 2000; Muchowska et al., 2017; Orgel, 2000).

A single-branching cycle has two key nodes: the branching reaction, where, in the autocatalytic direction, one member is a reactant and two members are products (Figure 1B, $R_j$); and the reunion member, which is a product of two reaction paths (Figure 1B, $M_1$). Thus, the topological structure of a simple autocatalytic cycle can be characterized by three parameters: $\rho$ (signifying "road"), the number of reactions between the reunion member and the reactant member of the branching reaction; $\pi_1$ (signifying "path 1"), the number of reactions between a product member of the branching reaction and the reunion member along the shorter branch; and $\pi_2$ (signifying "path 2"), the number of such reactions along the longer (or equal-length) branch. Table 1 shows the examples of 0-0-0, 1-0-0, 1-1-1, and 1-1-2 cycles. Note that the 0-0-0 cycle is a simple reversible autocatalytic reaction.

| Cycle name | Path type | Number of reactions | Reaction | Reunion member |
|---|---|---|---|---|
| 0-0-0 | $\rho$ | 0 | – | M |
| | Branching reaction | 1 | $M + F \leftrightarrow 2M$ | |
| | $\pi_1$ | 0 | – | |
| | $\pi_2$ | 0 | – | |
| 1-0-0 | $\rho$ | 1 | $M_1 + F \leftrightarrow M_2 + W$ | $M_1$ |
| | Branching reaction | 1 | $M_2 + F \leftrightarrow 2M_1$ | |
| | $\pi_1$ | 0 | – | |
| | $\pi_2$ | 0 | – | |
| 1-1-1 | $\rho$ | 1 | $M_1 + F \leftrightarrow M_2 + W$ | $M_1$ |
| | Branching reaction | 1 | $M_2 + F \leftrightarrow M_3 + M_4$ | |
| | $\pi_1$ | 1 | $M_3 + F \leftrightarrow M_1 + W$ | |
| | $\pi_2$ | 1 | $M_4 + F \leftrightarrow M_1 + W$ | |
| 1-1-2 | $\rho$ | 1 | $M_1 + F \leftrightarrow M_2 + W$ | $M_1$ |
| | Branching reaction | 1 | $M_2 + F \leftrightarrow M_3 + M_4$ | |
| | $\pi_1$ | 1 | $M_3 + F \leftrightarrow M_1 + W$ | |
| | $\pi_2$ | 2 | $M_4 + F \leftrightarrow M_5 + W$ <br> $M_5 + F \leftrightarrow M_1 + W$ | |

**Table 1. Examples of simple autocatalytic cycles.**

We model autocatalytic systems in a continuously stirred flow reactor (Figure 1C). The source solution of the food is constantly added into the inflow port (I) and the solution in the reactor is also constantly removed via the outflow port (O). The rates of addition, $k_v$, and removal are the same, meaning that $k_v$ is the dilution rate (Figure 1C), which is defined as the volume of the solution flowing into (or out from) the reactor per unit time divided by the volume of the solution in the reactor. To understand how such a reactor may model the prebiotic environment, we might view this reactor as a small, well-mixed pool in a hydrothermal field receiving a constant flux of liquid from an uphill volcano and simultaneously losing a similar amount through a downhill outlet (Damer and Deamer, 2015). Alternatively, we can view the reactor as a small patch of mineral at the bottom of the ocean, with chemicals held in the "reactor" by adsorption onto the mineral surface, and the "flow" representing adsorption/desorption occurring at the interface between the mineral and an ocean rich in food and poor in waste or members (Wächtershäuser, 1988).

In our models, the reactor has an additional entrance (P) for introducing additional chemicals, such as members, at different times. Such an entrance represents occasional introduction of new chemicals into the system. Scenarios where introduction might occur include rare geological events such as meteoritic impacts, long-distance import of chemicals from other chemical environments, and the occurrence of rare chemical reactions. In this paper, before adding members through this entrance, the concentrations of all input food in the reactor are set to be the same as that of the source solution. All chemicals in these models are assumed to be soluble at all concentrations considered.

## 2. The dynamics of a single autocatalytic cycle can be approximated by logistic growth model

In population ecology, the logistic growth model is widely used to describe how population size changes over time when resources are limited. The basic logistic growth model is formalized by the differential equation:

$$\frac{\mathrm{d}N}{\mathrm{d}t} = rN\left(\frac{K-N}{K}\right) \tag{1}$$

where $N$ is the population size, $t$ is time, $r$ is the intrinsic growth rate, and $K$ is the carrying capacity. This equation shows that $N$ will finally become stable near $K$, and that the speed with which $N$ approaches $K$ is governed by $r$.

In this section, we will show that the dynamics of a single autocatalytic cycle can be approximated by the logistic model. Furthermore, we determine that there are intrinsic connections between chemical kinetic parameters such as rate constants and ecological parameters such as the intrinsic growth rate.

### 3.1 The logistic growth of the 0-0-0 cycle

First, we analyze the 0-0-0 cycle:

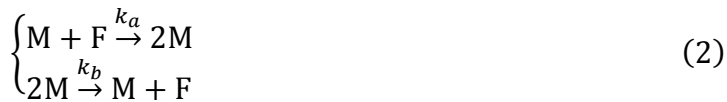

$$\begin{cases} M + F \xrightarrow{k_a} 2M \\ 2M \xrightarrow{k_b} M + F \end{cases} \tag{2}$$

where M is the sole member, F is the food, $k_a$ and $k_b$ are rate constants of the forward (autocatalytic) and reverse (reverse autocatalytic) reactions, respectively.

The concentration of food in the source solution is denoted by the constant $f$ and the dilution rate by $k_v$. It can be shown that if the initial concentration of M is much smaller than $f$, the dynamics of [M] (i.e., the concentration of M) can be approximated by the logistic model (see Supplemental Materials, Section 1):

$$\frac{\mathrm{d}[M]}{\mathrm{d}t} = r_M[M]\left(\frac{K_M - [M]}{K_M}\right) \tag{3}$$

by defining:

$$\begin{cases} r_M = k_a f - k_v \\ K_M = \dfrac{k_a f - k_v}{k_a + k_b} \end{cases} \tag{4}$$

This implies that, provided that $k_a f$ is greater than $k_v$, seeding the reactor with a tiny amount of M will result in logistic growth of [M] (Figure 2). This conclusion is not surprising and was already reported in the literature (Lloyd, 1967). Although the intrinsic growth rate and carrying capacity seem to be independent constants in the logistic equation (Equation (1)), we can see that $r_M$ and $K_M$ are actually non-independent because they are linked directly to the underlying rate constants. Lowering the rate constant of the reverse reaction, $k_b$, which amounts to making the net reaction more thermodynamically favorable, raises $K_M$ without altering $r_M$. In contrast, adding a catalyst, which raises both $k_a$ and $k_b$ without changing their ratio, increases both $r_M$ and $K_M$ (for details see Supplemental Materials Sections 2 and 3).

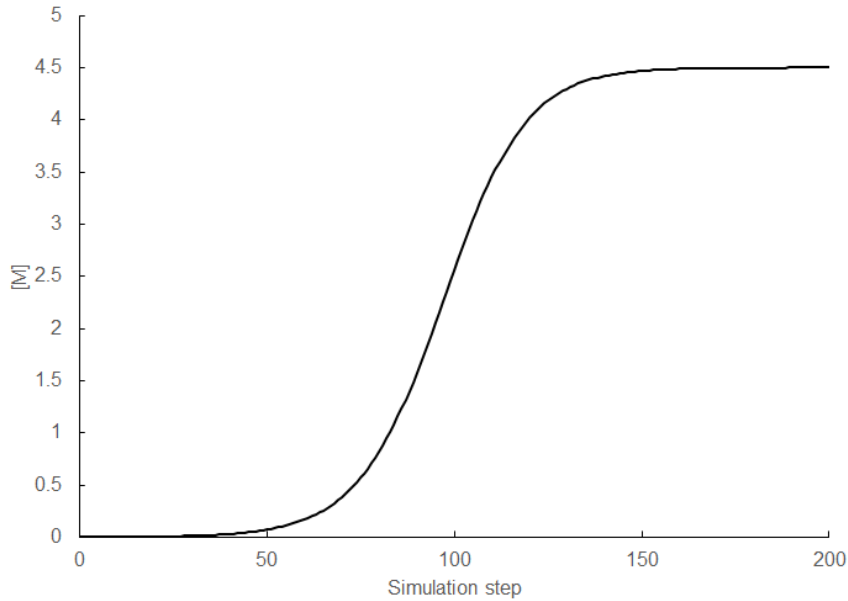

**Figure 2. The logistic growth of a 0-0-0 cycle**. This logistic growth curve of a 0-0-0 cycle was generated by setting $k_a = 0.01$, $k_b = 0.01$, $k_v = 0.01$, $f = 10$, and the initial concentration of M, $[M]_0 = 0.001$.

While increasing the dilution rate $k_v$ can increase the amount of food flowing into the reactor per unit time, it actually decreases $r_M$ and $K_M$ because it also increases the loss rate of the member. We can also see that, maximum growth rate and carrying capacity are reached when $k_v$ approaches zero, which makes the flow reactor become a closed reactor. This shows that there is a "cost" to openness: when the rate of member production and dilution balance out, there will be more unused food in the reactor than in the case that the system had reached equilibrium without flow. However, from a chemical point of view, openness seems inevitable and necessary. This is because: (a) all systems on the planet, and even the planet itself, are open systems receiving a flux of free energies in different forms; (b) only open systems have the potential for complex, long-term dynamical changes (Wagner et al., 2019); and (c) although an autocatalytic system may achieve maximum growth and system size locally in one closed reactor, its success would not be replicable in other environments. Indeed, if the member flowing out from a reactor can later flow into new reactors, increasing $k_v$ could be beneficial for the autocatalytic system to maximize global dispersal. Since the amount of M flowing out from a reactor per unit time is proportional to $k_v$[M], which can be defined as a dispersal index, it is easy to show that maximum dispersal will be achieved when $k_v / k_a = f / 2$ (see Supplemental Materials, Section 4).

### 3.2 The logistic growth of the 1-0-0 cycle

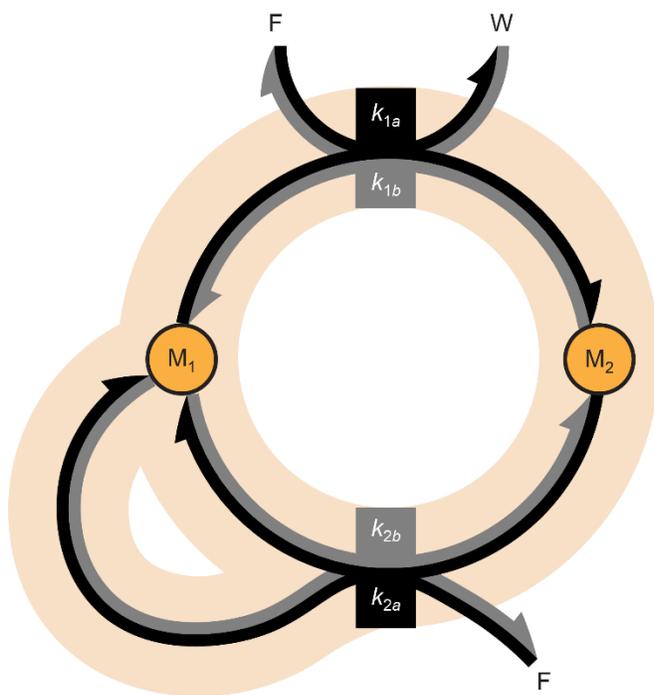

**Figure 3. The topological structure of the 1-0-0 cycle**. Black arrows point in the autocatalytic (forward) direction, whereas grey arrows point in the direction of reverse autocatalysis. The reaction boxes containing forward and reverse rate constants are color coded black and grey, respectively. The two forward arrows leaving the branching reaction (with the rate constant $k_{2a}$) indicate a stoichiometric doubling.

Now we proceed to the 1-0-0 cycle (Figure 3). By making the substitutions:

$$\begin{cases} S = [M_1] + [M_2] + [W] \\ [M_1] = \theta_1 S \\ [M_2] = \theta_2 S \\ [W] = (1 - \theta_1 - \theta_2) S \end{cases} \tag{5}$$

it can be shown that the dynamics of $S$ can be approximated by the logistic model (see Supplemental Materials, Section 5):

$$\frac{\mathrm{d}S}{\mathrm{d}t} = r_S S \left( \frac{K_S - S}{K_S} \right) \tag{6}$$

by defining:

$$\begin{cases} r_S = k_{1a} \theta_1 f + k_{2a} \theta_2 f - k_v \\ K_S = \dfrac{k_{1a} \theta_1 f + k_{2a} \theta_2 f - k_v}{k_{1a} \theta_1 + k_{2a} \theta_2 + k_{1b} \theta_2 (1 - \theta_1 - \theta_2) + k_{2b} {\theta_1}^2} \end{cases} \tag{7}$$

Because $\theta_1$ and $\theta_2$ are not necessarily constants, $r_S$ and $K_S$ may vary with $\theta_1$ and $\theta_2$ across different stages of the growth dynamics. This suggests that the $[M_1]$-to-$[M_2]$ ratio of the initially added members may have a significant impact on the growth dynamics, at least initially. On the other hand, numerical simulations show that $\theta_1$ and $\theta_2$ are nearly constant after the early stage of growth (Figure 4), so we can approximately treat $r_S$, $K_S$, $\theta_1$, and $\theta_2$ as constants. Thus, according to Equations (6) and (7), the growth dynamics of $[M_1]$, $[M_2]$, and $[W]$ can each also be approximated by logistic models (Figure 4). In addition, the fact that $r_S$ and $K_S$ are related to $\theta$ values shows that the growth dynamics of an autocatalytic cycle will be affected by the choice of members to seed the cycle. Indeed, numerical simulations show that the seed choice does impact the initial growth rate, but without affecting the carrying capacity (see Supplemental Materials, Section 6). The initial growth rate is the highest if the seed is the reactant member of the branching reaction, and the lowest if it is the reactant of the first reaction on the $\pi_2$ branch (Figure S2).

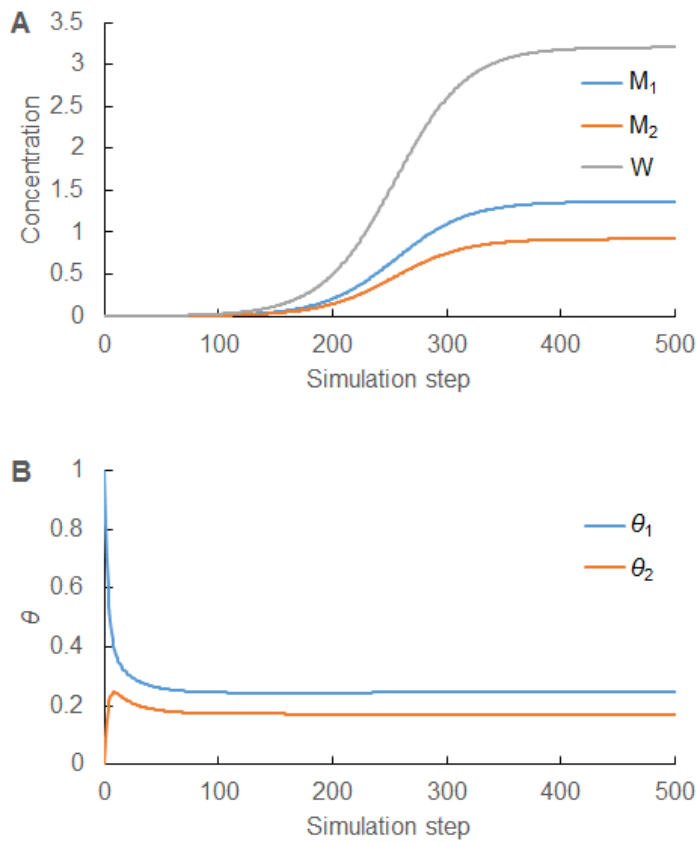

**Figure 4. Logistic growth dynamics of the 1-0-0 cycle.** **(A)** The dynamics of waste and members. **(B)** The dynamics of $\theta$ values. $\theta_j$ is the proportion of member $M_j$ in the total concentration of waste and all members. This simulation was run by setting all rate constants to 0.01, $f = 10$, $k_v = 0.01$, $[M_1]_0 = 0.001$ and $[M_2]_0 = 0$.

Following the same logic, the growth dynamics of other autocatalytic cycles of this basic form in a flow reactor are likely to be approximated by logistic models provided that the starting concentrations of members are much smaller than the food concentration, and/or the starting ratios of the members' concentrations are close to the ratio that they will attain at steady state. This general logistic form supports a general analogy between an autocatalytic cycle and a population of organisms. To link the two together, we may imagine that the chemical cycle represents the life cycle of a single individual. For example, a 1-0-0 cycle can be seen as an egg becoming an adult and then an adult producing two eggs and dying, with food needed to convert an egg into an adult and an adult into two new eggs. There are, of course, major differences between autocatalytic chemical cycles and biological organisms, most notably the reversibility of the former. You cannot feed organisms waste and see them shrink while spitting out food!

## 4. Numerical analyses on behavior of more complicated cycles

Although analytical treatment can generate exact descriptions of the growth dynamics, it is not always feasible to obtain such solutions, especially when the autocatalytic cycle consists of more chemicals and reactions. Thus, hereafter, we use numerical simulations to explore factors that affect the dynamics of more complicated cycles and chemical ecosystems containing more than one autocatalytic cycle. First, we will use numerical analyses to examine how chemical kinetic parameters affect the growth dynamics of single autocatalytic cycles.

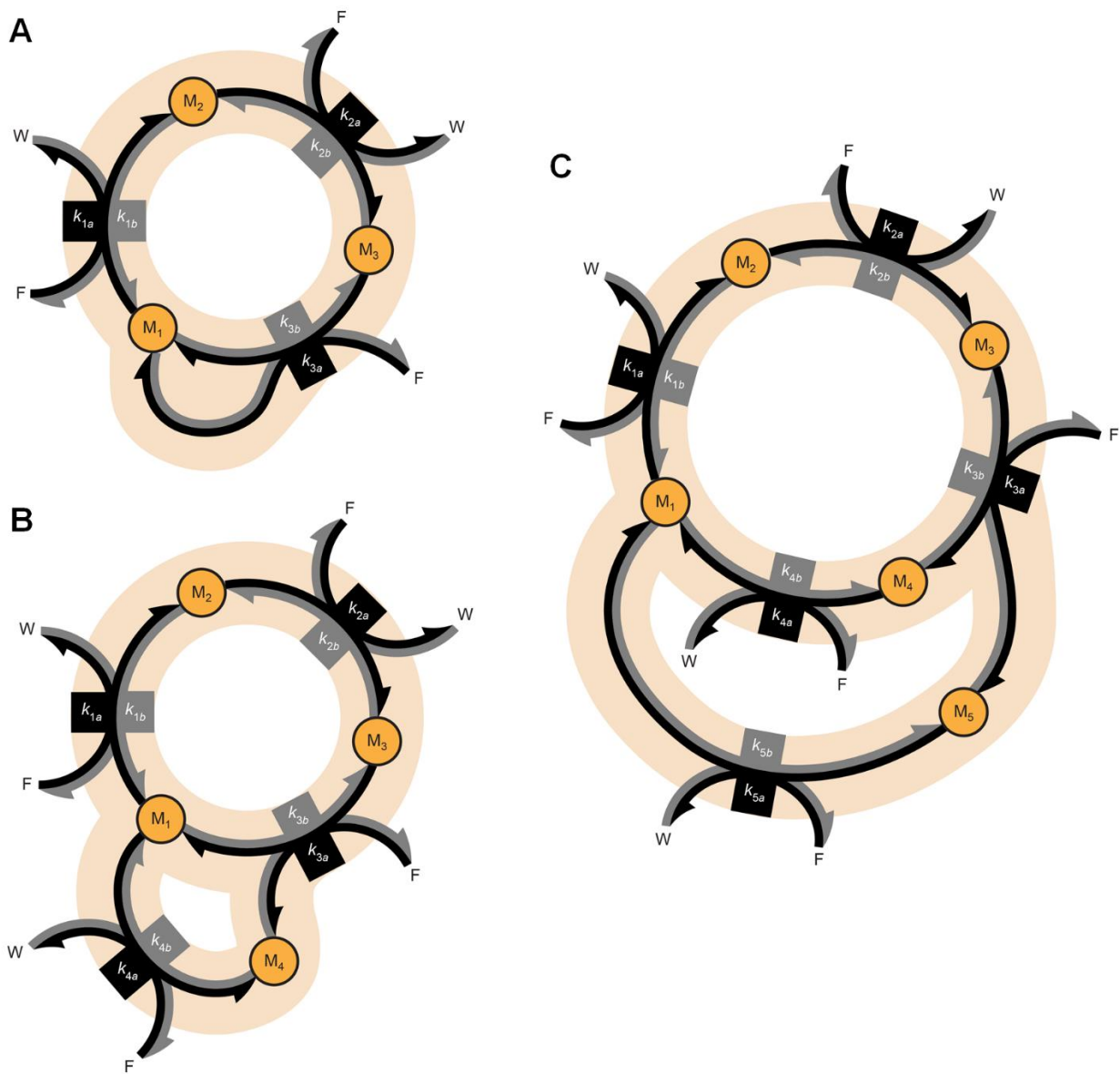

**Figure 5. The topological structures of the 2-0-0, 2-0-1, and 2-1-1 cycles.** (**A**) The 2-0-0 cycle. (**B**) The 2-0-1 cycle. (**C**) The 2-1-1 cycle. The arrows link chemicals, which are either members (M), food (F), or waste (W). Black arrows point in the autocatalytic (forward) direction, whereas grey arrows point in the reverse direction. The reaction boxes containing forward and reverse rate constants are color coded black and grey, respectively.

We considered the 2-0-0, 2-0-1, and 2-1-1 cycles (Figure 5) and ran multiple simulations with the same (symmetrical) rate constants (0.01), but different $k_v$ and $f$. This allowed us to explore the relationship between the parameters determining food flux, namely the dilution rate $k_v$ and the input food concentration $f$, and the parameters describing growth (the intrinsic growth rate $r_M$, the carrying capacity $K_M$) and maximal potential dispersal ($k_v K_M$). We explored $k_v \in [0, 0.012]$ and $f \in [0, 4]$ (Figure 6).

Not surprisingly, for the same combinations of $f$ and $k_v$, networks with more members have lower values of $r_M$ (Figure 6A, D, G), $K_M$ (Figure 6B, E, H), and $k_v K_M$ (Figure 6C, F, I). This follows since a larger cycle requires more food for each iteration of the full cycle. It is also intuitive that larger cycles have a higher threshold flux rate ($f / k_v$) needed for the cycle to grow and sustain itself (i.e., $K_M$ exceeding the tiny initial concentration of members). In addition, if we define the threshold of $f / k_v$ as $\eta$, then for $f / k_v > \eta$, $r_M$ and $K_M$ are generally positively correlated to $(f - \eta k_v)/\sqrt{1 + \eta^2}$ (see Supplemental Materials, Section 7). Concerning the potential dispersal, for a given $f$, the maximal $k_v K_M$ is achieved when $k_v$ is approximately $f / (2\eta)$ (Figure 6C, F, I), which is consistent with the well-known principle that the maximum sustainable yield of a logistically growing population is achieved when the population size is half its carrying capacity.

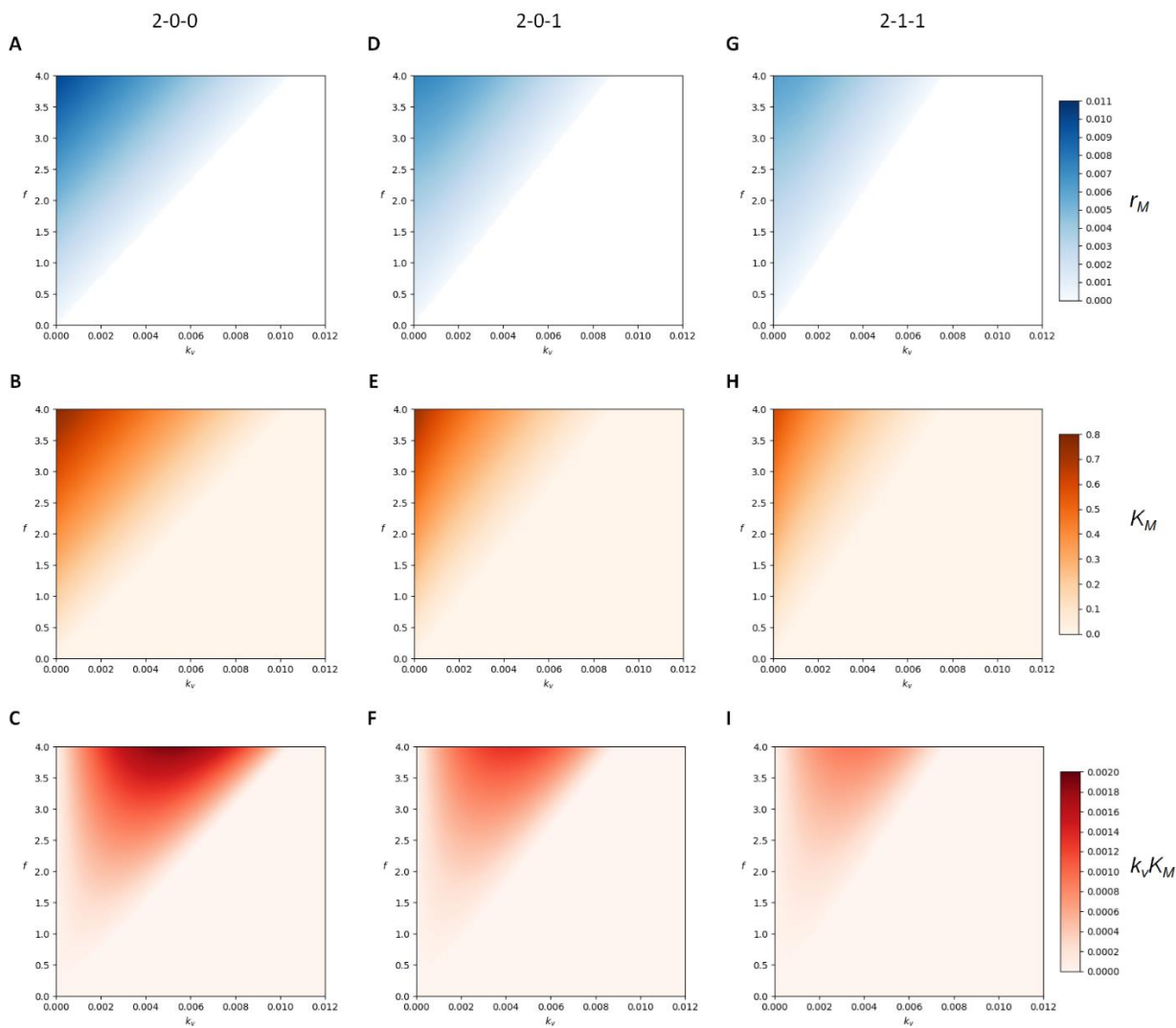

**Figure 6**. **The dilution rate, $k_v$, and food concentration in the source solution, $f$, affect the growth rate ($r_M$), carrying capacity ($K_M$) and potential dispersal ($k_v K_M$) of autocatalytic cycles**. **(A)** - **(C)** 2-0-0 cycle. **(D)** - **(F)** 2-0-1 cycle. **(G)** - **(I)** 2-1-1 cycle. The simulations were run by setting all rate constants to 0.01 and $[M_1]_0 = 10^{-6}$, with all other members at 0 concentration. For each combination of $k_v$ and $f$, the simulation was run to estimate $K_M$ and $\max(\mathrm{d}[M] / \mathrm{d}t)$, where $[M]$ is the sum of concentrations of all members. Then $r_M$ was calculated according to $K_M$ and $\max(\mathrm{d}[M] / \mathrm{d}t)$. Note that for **A**, **D**, and **G**, non-positive $r_M$ values are not shown.

## 5. Chemical interactions between autocatalytic cycles mimic ecological interactions between biological species

The prior sections showed that an autocatalytic cycle in a flow reactor has many similarities to a population of organisms. This suggests that if multiple autocatalytic cycles are allowed to chemically interact within a flow reactor, their dynamics might be similar to those of ecological interactions between populations. To investigate this explicitly, we examined different types of interactions between pairs of autocatalytic cycles. The scenarios examined below are only a subset of the possible interactions among autocatalytic cycles.

## 5.1 Competition

In ecology, competition arises when multiple species compete for the same resources and those resources are limited (Neill et al., 2009; Passarge et al., 2006; Sommer, 1999). For example, plants compete for sunlight and predators compete for prey. In ecology, in cases where at least one of the species is viable in a certain ecosystem, possible results of pairwise competition are: (a) competitive coexistence, where competitors exist in the same environment and (b) competitive exclusion, where one competitor survives and one goes extinct. Using two simple 0-0-0 cycles that share the same food (Figure 7A), it can be shown that there are three possible steady states (see Supplemental Materials , Section 8), one corresponding to coexistence and two to exclusion, as follows.

$$\begin{cases} [\text{M}] = \dfrac{k_a k_\beta f - k_v(k_\alpha + k_\beta - k_a)}{k_a k_\beta + k_\alpha k_b + k_b k_\beta} \\ [\mu] = \dfrac{k_\alpha k_b f - k_v(k_a + k_b - k_\alpha)}{k_a k_\beta + k_\alpha k_b + k_b k_\beta} \end{cases} \tag{8}$$

$$\begin{cases} [\text{M}] = \dfrac{k_a f - k_v}{k_a + k_b} \\ [\mu] = 0 \end{cases} \tag{9}$$

$$\begin{cases} [\text{M}] = 0 \\ [\mu] = \dfrac{k_\alpha f - k_v}{k_\alpha + k_\beta} \end{cases} \tag{10}$$

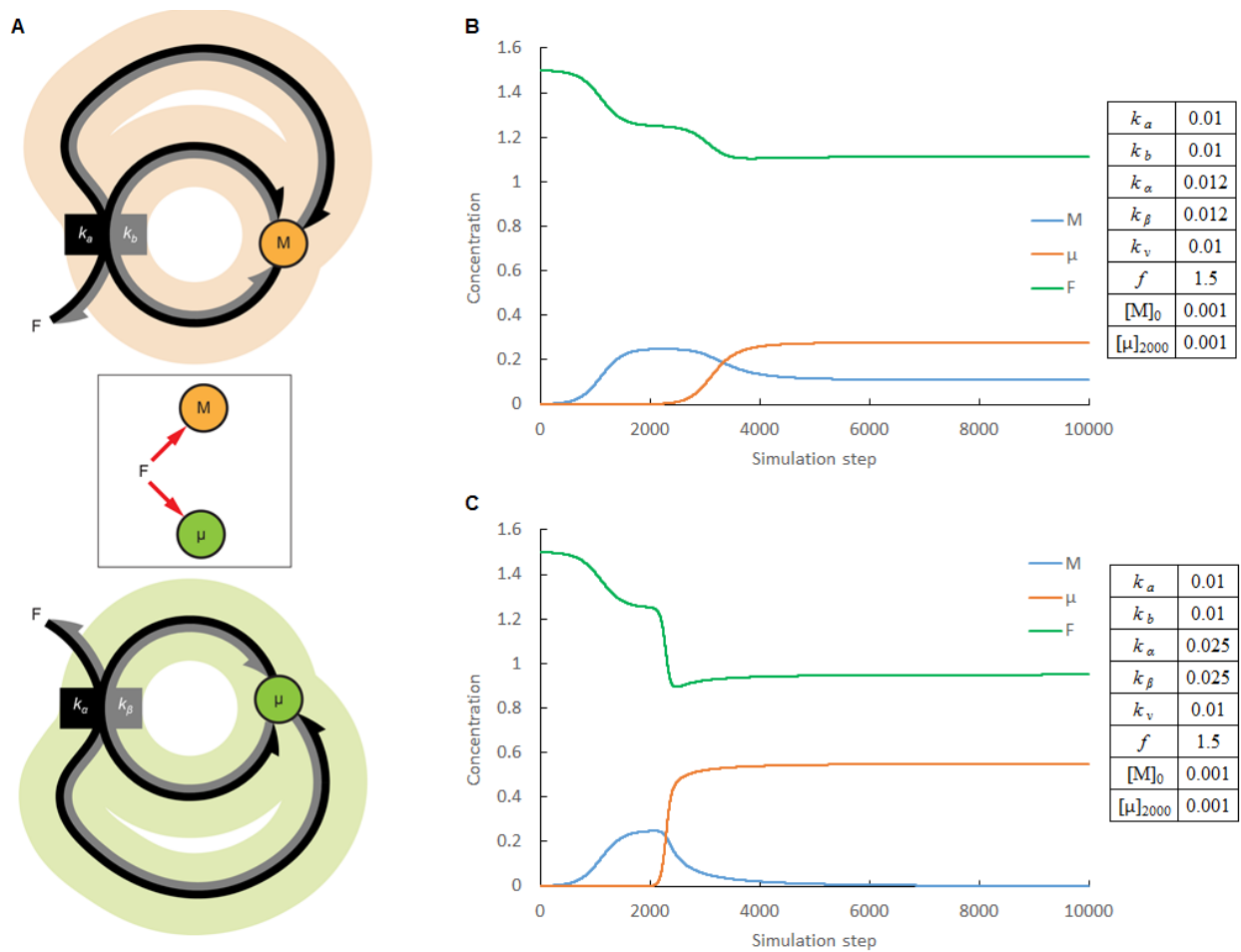

**Figure 7. Competition**. The results of the competition between two 0-0-0 cycles in a flow reactor. (**A**) M (shown by the orange shade) and μ (shown by the green shade) compete for the same food F. The subpanel summarizes their interaction. (**B**) Competitive coexistence. (**C**) Competitive exclusion. Parameters used to generate the dynamics are shown next to the dynamics. In both cases, M was introduced at the beginning and μ was introduced at the 2000th step.

To visually present the dynamics of competition, we numerically simulated two representative scenarios (Figure 7). For both scenarios, the M cycle is assumed to have lower rate constants than the μ cycle, making μ a better competitor. In each scenario, M was introduced at the beginning, with μ introduced later. When the rate constants are close in magnitude, competitive coexistence occurs, where μ suppresses M but M can still persist (Figure 7B). In contrast, when the competitive ability of μ relative to M exceeds a threshold, competitive exclusion occurs, where the introduction of μ results in M declining towards zero concentration. It is worth noting that for the scenarios shown in Figure 7, the ratio between the rate constants of the forward and reverse reactions is kept constant. Therefore, these results illustrate that catalyzed reactions in an autocatalytic cycle increase the competitiveness of that cycle, despite the fact that catalysis accelerates both the forward and reverse reactions. This implies that cycles that happen to be able to utilize environmental chemicals (e.g., metal ions, protons) as their catalysts, are more likely to persist than equivalent cycles that do not have such an ability. In addition, if the reaction network is more complex such that the members or waste of a cycle can be converted into the catalysts of the cycle, the network as a whole is likely to have higher competitiveness.

It is worth noting that when an ecosystem is taken over by a better competitor, the efficiency with which the food flux is exploited generally increases, as seen by a significant decline in the concentration of food in the reactor (Figure 7B-C). However, it should be noted that such increase in the efficiency is not strictly monotonic, as the energies stored in M would be gradually released as the food while μ is establishing its dominance (Figure 7B-C). The increase is more profound when we compare between the M-dominating and the μ-dominating stages. This confirms the intuition that chemical ecosystems are dissipative systems driven by food and that, as a result, there is a tendency for these systems to become progressively better at dissipating this energy (Baum, 2018).

In the case of competition between larger autocatalytic cycles, it is possible for competition to be for food, or waste, or both. This follows because the accumulation of waste can facilitate reverse autocatalysis, which means that for an autocatalytic cycle, if there are other competitor cycles generating the same waste, its growth will be suppressed. Numerical simulations (not shown) confirmed that the range of possible outcomes is the same when competition is for waste rather than food.

## 5.2 Mutualism

Since metabolic networks of living organisms consist of a large number of chemicals and reactions that appear to cooperate, "cooperation" between chemicals and reactions has long been considered as an important factor in the origin of life (Ehrenfreund et al., 2006; Herschy et al., 2014; Lanier et al., 2017; Mathis et al., 2017; Pereira et al., 2012; Vaidya et al., 2012). Before the origin of metabolic control via the encoded production of specific catalysts, cofactors, and inhibitors, mutualistic interactions among autocatalytic systems could explain the emergence of metabolic complexity. In facultative mutualism, different species can survive without each other but nonetheless gain a benefit from each other's presence. For example, omnivores benefit from fruits but can also rely on other food sources, and plants benefit from omnivores spreading the seeds but may also disperse seeds by gravity. In obligate mutualism, cooperating species require one another for survival or reproduction. For example, figs and fig wasps form reciprocally obligate mutualism, as figs require fig wasps to reproduce and *vice versa*. In this section, we will show how the interactions between two 2-0-0 cycles can generate the dynamics of facultative mutualism, unidirectionally obligate mutualism, and reciprocally obligate mutualism.

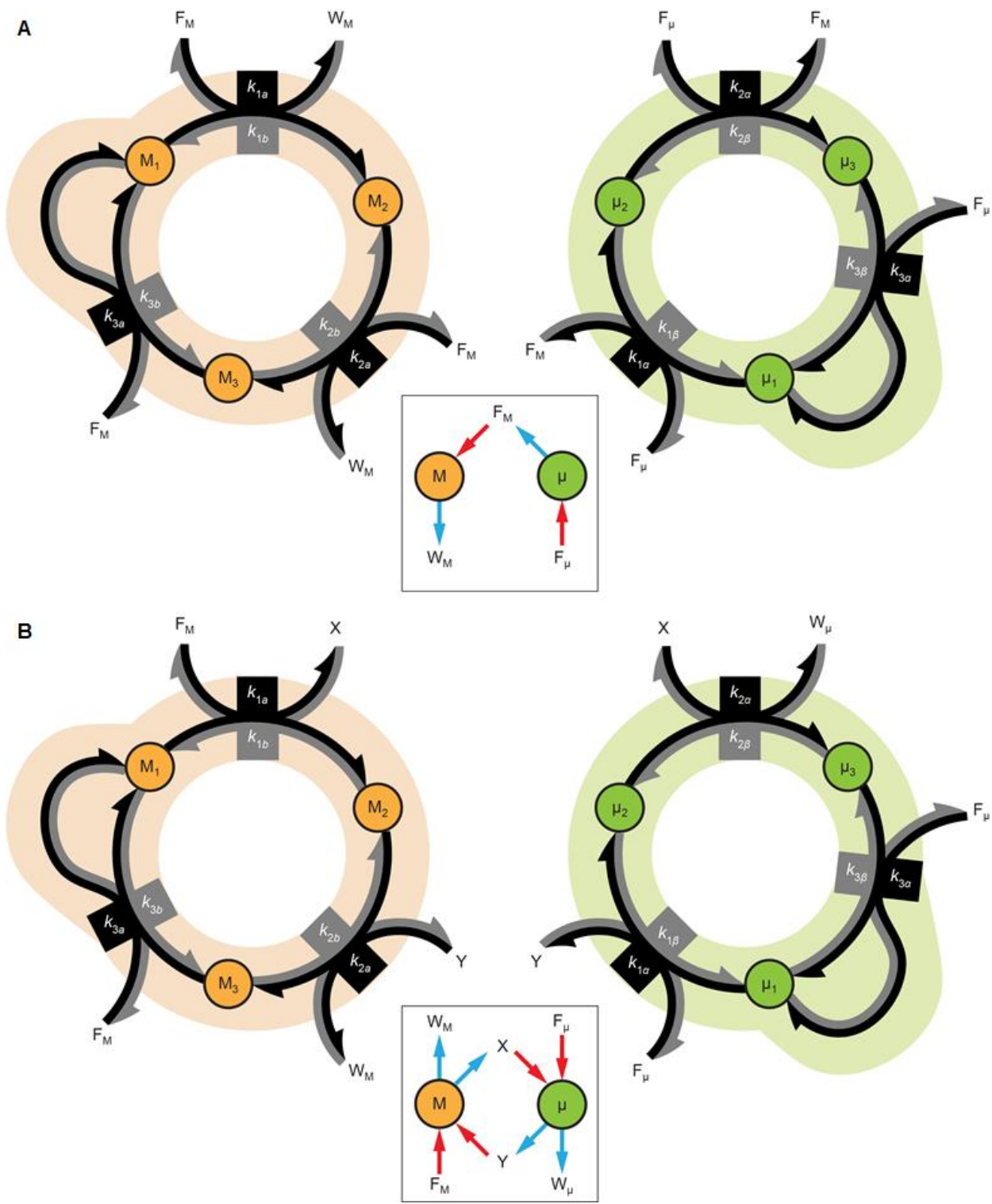

**Figure 8. Possible reaction networks underlying mutualism. (A)** Facultative mutualism and unidirectionally obligate mutualism. **(B)** Reciprocally obligate mutualism.

Facultative and obligate mutualisms can be generated between the cycles where the food of one is the waste of the other (Figure 8). In our model of facultative mutualism (Figure 8A, Figure 9A-C), when reactions are initiated just by $M_1$ (Figure 9A) or $\mu_1$ (Figure 9B), the M cycle or $\mu$ cycle, respectively, can survive in the reactor. However, the carrying capacities of the M and $\mu$ cycles are higher when the reactions are initiated by both $M_1$ *and* $\mu_1$ (Figure 9C). This is because the production of $F_M$ by the $\mu$ cycle provides additional food for the M cycle and the consumption of $F_M$ by the M cycle helps remove waste from the $\mu$ cycle, promoting reaction in the autocatalytic direction. Therefore, the interaction between these two cycles is a facultative mutualism.

To model unidirectionally obligate mutualism, the only modification needed is to no longer add $F_M$ through the entrance I (Figure 1C) such that $f_M = 0$. Thus, the M cycle completely relies on the $F_M$ produced by the $\mu$ cycle, whereas the $\mu$ cycle can survive without the M cycle. In this case, if the reactions are initiated by both $M_1$ and $\mu_1$, the M cycle can survive as the $\mu$ cycle produces $F_M$ (Figure 9D), and the $\mu$ cycle can have a higher carrying capacity as its waste is consumed by the M cycle (Figure 9D). In this scenario, the presence of the $\mu$ cycle is necessary for the survival of the M cycle. However, the presence of $\mu$ is not sufficient to guarantee that M persists: $f_\mu$ could be low enough that it can support the growth of $\mu$ but results in the production of too little $F_M$ to support the growth of M.

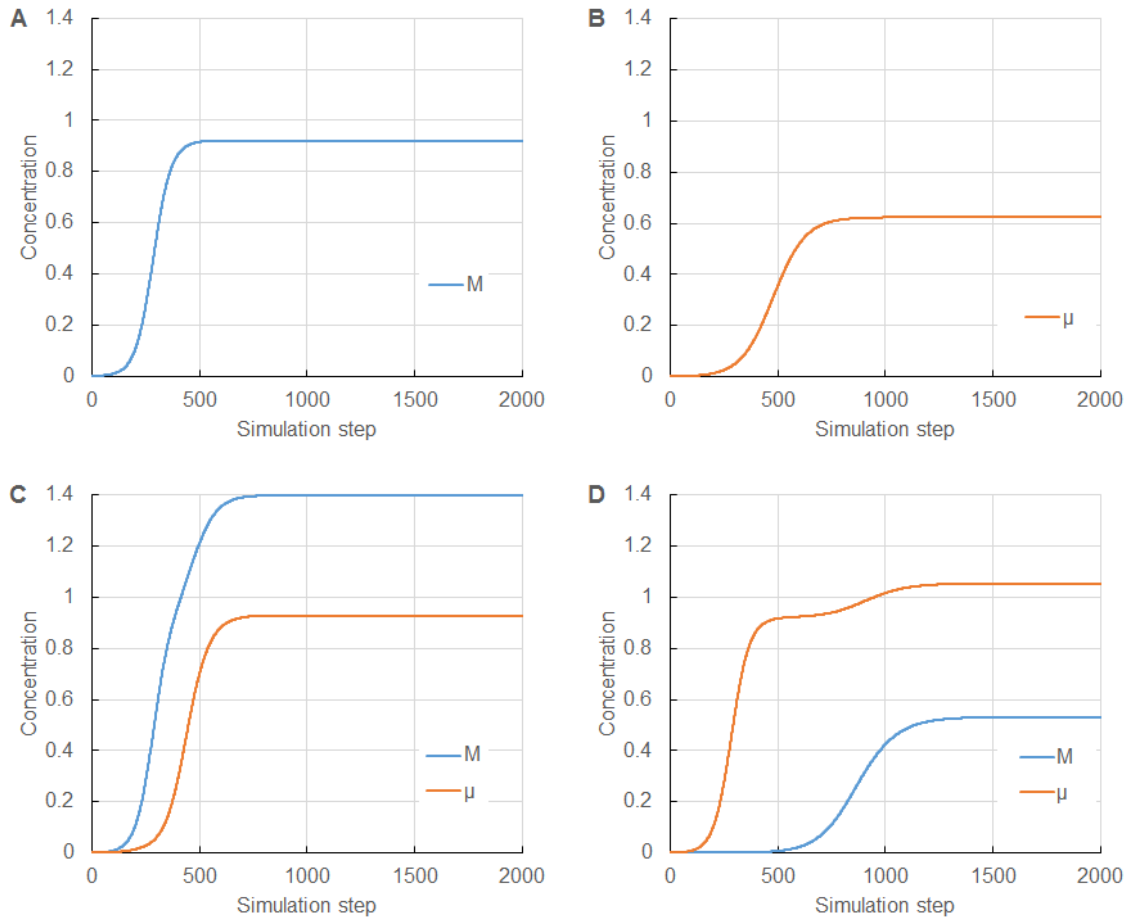

**Figure 9. Facultative and unidirectionally obligate mutualisms due to recycled waste**. The vertical axis shows the total concentration of members for the M and/or μ cycles. All rate constants are 0.02, and $k_v = 0.001$. (**A-C**) correspond to facultative mutualism because food for both cycles is provided ($f_M = f_\mu = 5$), whereas (**D**) corresponds to unidirectionally obligate mutualism because only $F_\mu$ is provided ($f_M = 0; f_\mu = 5$). Simulations **A-C** differ in whether they are seeded with M only (**A:** $[M_1]_0 = 0.001$), μ only (**B:** $[\mu_1]_0 = 0.001$), or both (**C:** $[M_1]_0 = [\mu_1]_0 = 0.001$).

To increase the benefits of cooperation still further, we can consider a case where the waste of one cycle is an indispensable food (i.e., one not provided from the external sources) for the other, and *vice versa* (Figure 8B). Such a strong, reciprocally obligate mutualism ties the two cycle's fate together closely (Figure 10). In these conditions, there is a long waiting time before the onset of fast growth of [M] and [μ]. Because of this interdependency, the growth curve differs from a typical logistic curve (Figure 10).

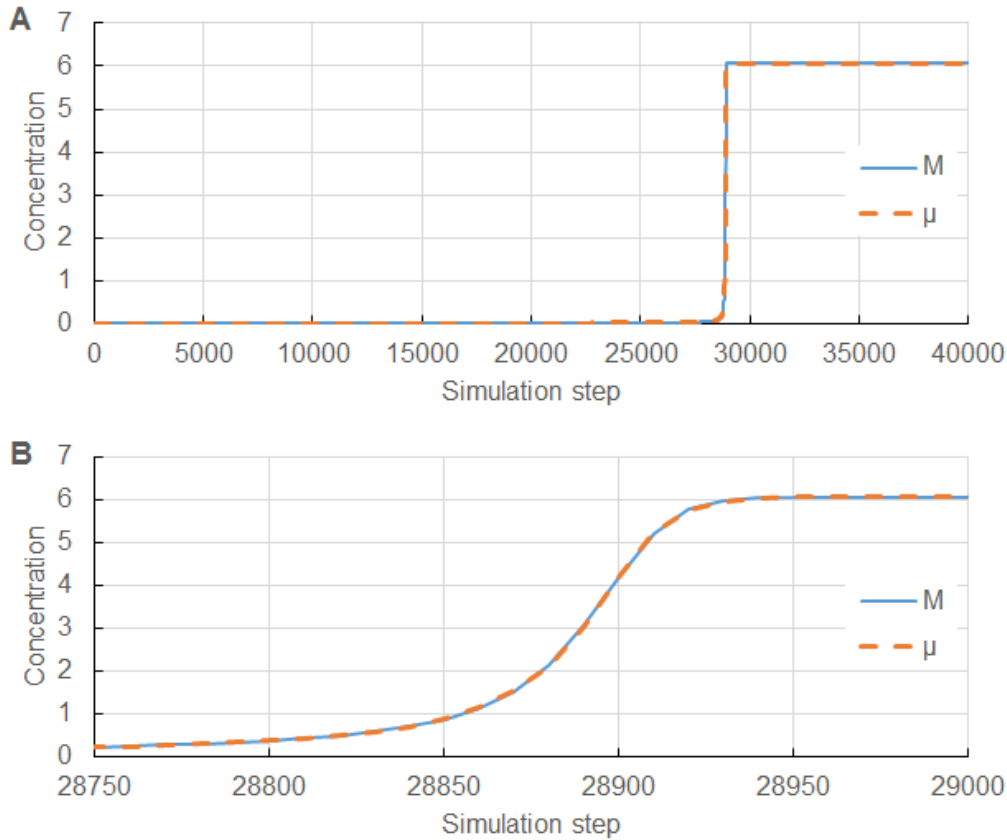

**Figure 10. Reciprocally obligate mutualism due to recycled waste**. (**A**) The dynamics of the total concentration of members for the M and μ cycles. (**B**) A zoomed-in view of the growth phase of the dynamics shown in **A** The simulation was run by setting all rate constants to 0.04, $k_v = 10^{-5}$, $f_M = f_\mu = 20$, $[M_1]_0 = [\mu_1]_0 = 0.001$.

## 5.3 Predation

Predation, in which predators kill and feed on prey species, is a common way for many organisms to obtain food. Pathogenesis and parasitism are similar to predation; pathogens and parasites also feed on host species but usually consume only part of the host, not killing the host immediately. Classically, the population dynamics of predator-prey systems is modelled using the Lotka-Volterra equations (Lotka, 1927, 1920; Volterra, 1927, 1926), which allow that predation may lead to diverse dynamics, including stable oscillation, damped oscillation, and establishment of a steady state without oscillation. In this section, we will show that if the members of one 2-0-0 cycle are the food of another (Figure 11), the dynamics will be similar to the Lotka-Volterra model.

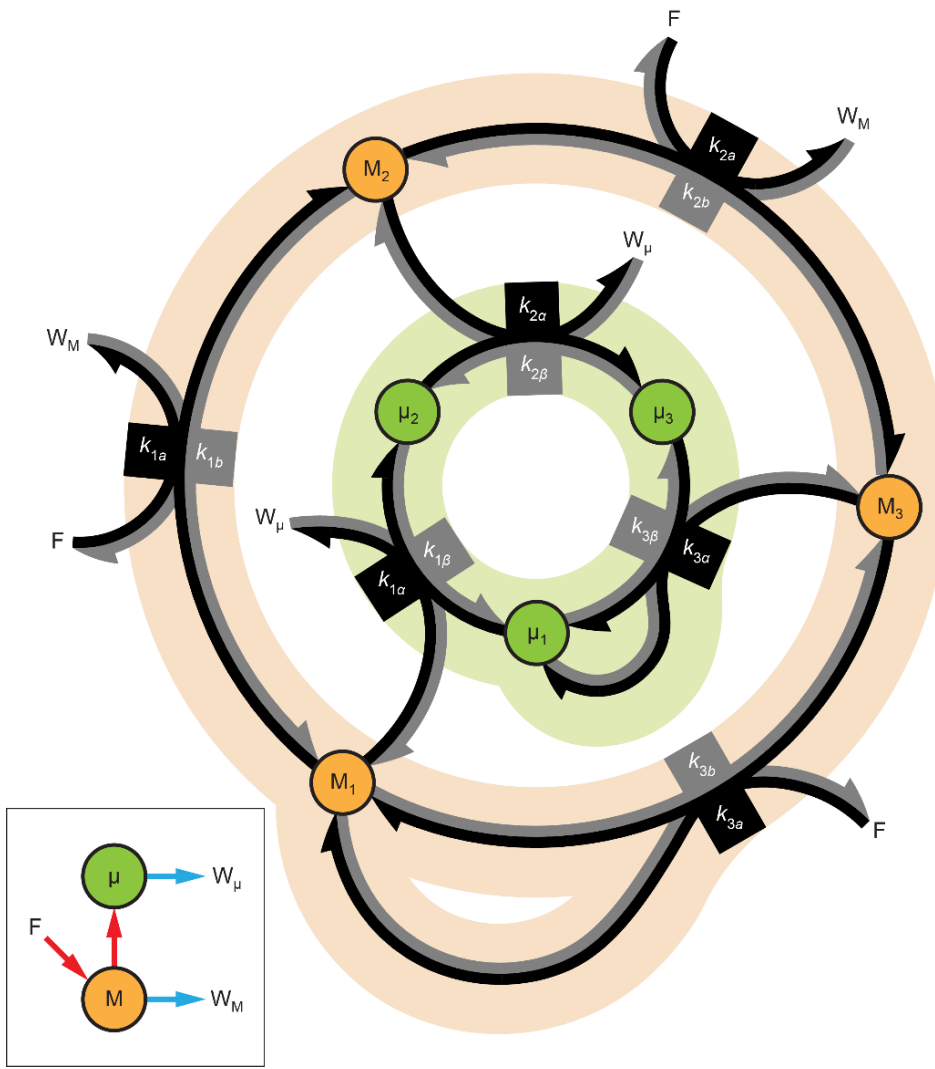

**Figure 11. A possible reaction network underlying predation.** The M cycle is the "prey" and the μ cycle is the "predator".

In numerical simulations, the parameters of the prey cycle, M, (Figure 11) were held constant, with equal forward and reverse rate constants, and we examined how the dynamics varied with different rate constants for the predator cycle, μ (Figures 11-12). If the predator rate constants are the same as the prey rate constants, the predator cycle cannot survive in the reactor (Figure 12A). If the predator rate constants are larger but equal in the forward (autocatalytic) and reverse directions, the predator cycle can survive in the reactor at steady state (Figure 12B). When the predator's rate constants for autocatalysis exceeds those of reverse autocatalysis, resulting in what we might think of as a higher predation efficiency, the dynamics can display damped (Figure 12C) or stable oscillations (Figure 12D). However, higher predation efficiency decreases the concentrations of both the predators and preys in the reactor (Figure 12B-D).

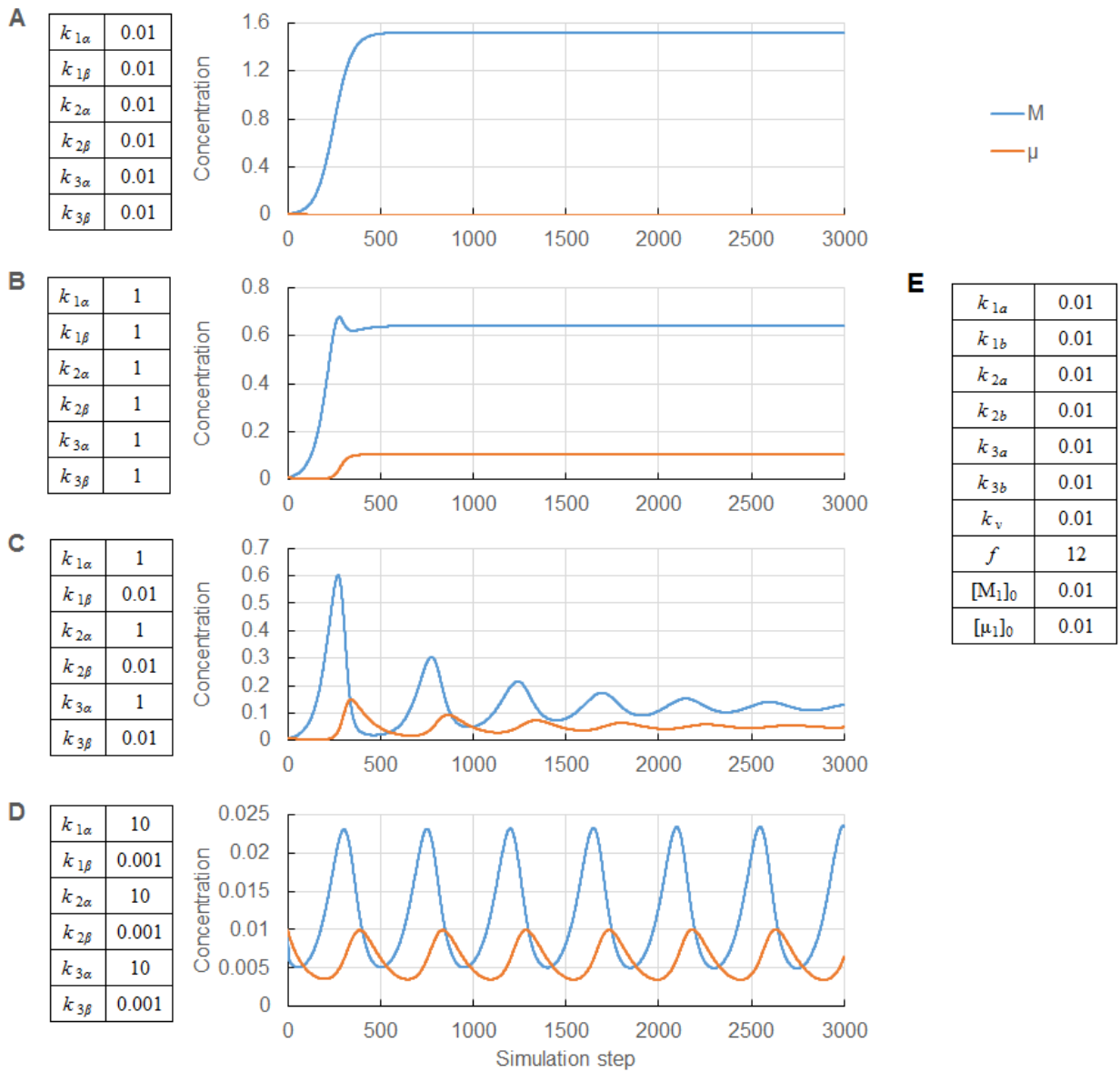

**Figure 12. Predation**. The dynamics of the total concentration of members for the M and μ cycles, when the M cycle is the prey cycle and the μ cycle is the predator cycle. The parameters specific to a simulation are shown to the left of panels **A-D**, while panel **E** shows parameters shared by all simulations. (**A**) Extinction of the predator cycle. (**B**) Steady state coexistence of the prey and predator cycle. (**C**) Damped oscillation of the prey and predator cycles. (**D**) Stable oscillation of the prey and predator cycles.

## 6. Ecosystem dynamics and the transition to evolution

We have shown that individual autocatalytic cycles show behavior similar to populations of individual species in an ecosystem and that pairs of cycles can exhibit interactions similar to those previously studied in ecology. How might our understanding of the origins of life, and especially the origins of adaptive evolution, be clarified by such a chemical ecosystems perspective? We will attempt to show here that the dynamical behavior of even simple ecosystems can come to resemble evolution in two ways. First, long term dynamical patterns can be observed when the internal dynamics are combined in specific ways. Second, the order in which different potential autocatalytic cycles become actualized can be sensitive to the history of the appearance of seeding members. We will not attempt, here, to extend the model to include transitional steps towards more familiar genetics-based evolution, though we believe that this is quite feasible within a chemical ecosystem ecology framework.

### 6.1 Succession

In ecology, succession refers to changes in species composition within a geographical range over time (Connell and Slatyer, 1977; Morin, 2011, pp. 319–339; Prach and Walker, 2011). Although such changes are gradual, the entire process can sometimes be separated into distinct stages, each with different dominating species (Morin, 2011, pp. 319–339). Insofar as these changes are autogenic, driven by the properties of the species in the ecosystem rather than changes in the external environment itself, which is commonly the case (Connell and Slatyer, 1977), the changes have an evolutionary character. The current state of an ecosystem in a particular area is a heritable phenotype because that state is self-sustained for at least a period of time. For example, an area dominated by grasses this year is likely to also be dominated by grasses next year. And even if the current state would change to a different one, the changes are usually gradual under a small timescale. For example, an area completely dominated by grasses this year may be mostly dominated by grasses while some shrubs may appear next year. While not sufficient for adaptive evolution, the heritability of a successional stage, rather similar to the epigenetic maintenance of cell type during the development of multicellular organisms, provides the first necessary ingredient.

Successional dynamics have two key components: a delay in the invasion of late successional species and (perhaps) the eventual extinction of early successional species. The former feature could involve early successional species facilitating the later establishment of late successional species. Of these potential mechanisms, we have already described some of the possible mechanisms. Trophic level climbing applies when the establishment of an early successional prey cycle/species can support the later establishment of a predatory cycle/species. Asymmetric mutualism can also explain this pattern when an early successional cycle/species produces waste that serves as essential food for a late successional cycle/species. And, though we have not modelled it here, it should be obvious that niche amelioration can occur, in which an early successional cycle/species removes an inhibitory factor from the environment that would otherwise make the late successional cycle/species inviable.

The mechanisms of autogenic succession just described all apply in both biological and chemical ecology, but it is worth considering a variant mechanism that only applies in the chemical case: successional seeding. This occurs when a member of an early successional cycle, namely one whose seed member already exists in the environment, can be converted (perhaps slowly) into a member of the late successional cycle, thereby actualizing the late successional cycle. It is as though an early successional species could mutate to produce a seed of a different species, something that is impossible in modern biology but possible in chemistry. For example, if all members are chiral molecules, the early successional cycle may consist of left-handed members while the late succession cycle may consist of right-handed members, and the conversion from left-handed molecules to right-handed molecules could be mediated by chemical reactions.

To explore this phenomenon, we used numerical simulations of the interaction between two 2-0-0 cycles in which a member of one cycle (denoted M) can be converted by a side reaction into a member of the other (denoted μ) (Figure 13). To include the tendency for late successional cycle to outcompete early successional cycle, this model assumes that the M and μ cycles compete for the same food and waste, with the μ cycle having higher rate constants.

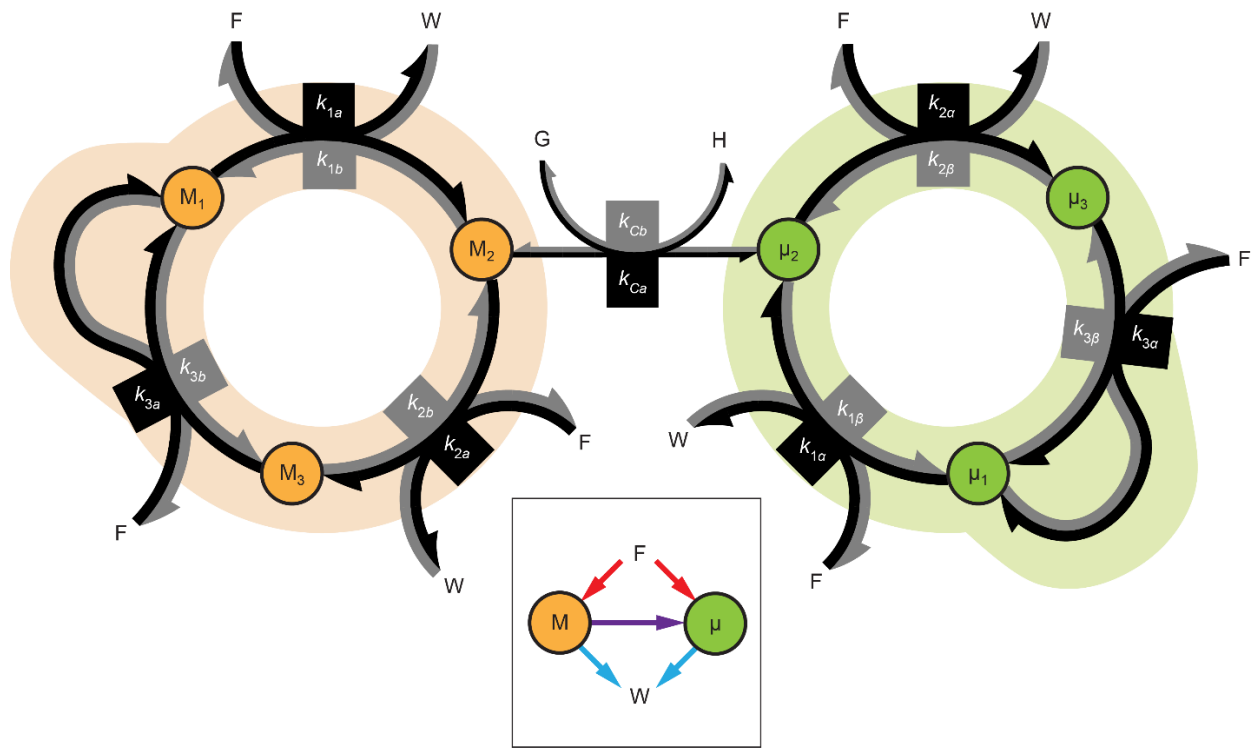

**Figure 13. A possible reaction network underlying succession by successional seeding.** The M cycle (shown by the orange shade) and the μ cycle (shown by the green shade) are linked by a reversible reaction $M_2 + G \leftrightarrow \mu_2 + H$, indicated by the thin arrows in the main panel. The subpanel summarizes the interactions between the two cycles. Note that the interconversion reaction is not a component of either autocatalytic cycle.

In this model, the cycles are linked by reversible reactions with additional chemicals H and G, as follows.

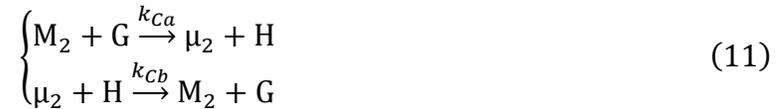

$$\begin{cases} M_2 + G \xrightarrow{k_{Ca}} \mu_2 + H \\ \mu_2 + H \xrightarrow{k_{Cb}} M_2 + G \end{cases} \qquad (11)$$

where $k_{Ca}$ and $k_{Cb}$ are rate constants. We assume that $k_{Ca}$ and $k_{Cb}$ are very low, representing slow reactions. We also assume that except for F and W, the chemicals $M_1$, G, and H are relatively simple whereas all other chemicals are too complex to emerge spontaneously, such that only $M_1$, G, and H may be introduced from outside the reactor. Two model variants are worth considering, deterministic and stochastic. In the deterministic model, G and H are included in the food solution at concentrations, $g$ and $h$, that are very low. In the stochastic model, G and H are injected stochastically through the entrance P (Figure 1C). The probability of injecting G per simulation step is $P_G$, and that of H is $P_H$. $P_G$ and $P_H$ are assumed to be very low (and equal), representing rare events. Each injection of G or H is assumed to increase [G] by $I_G$ or [H] by $I_H$, respectively. This mimics a case in which the early successional species are able to establish and remain dominant because propagules of late successional (invasive) species are rare.

In the deterministic model, in which G and H are added throughout the run (Figure 14), there is a long waiting time followed by a relatively fast transition from an M-dominated stage to an μ-dominated stage. When G and H are added stochastically (Figure 15), there is also a long waiting time between the first addition of G and the transition from M- to μ-domination (Figure 15A-B). Not shown here, but we can also observe succession-like dynamics if the waste of one autocatalytic cycle is slowly and/or rarely converted to the members of another cycle.

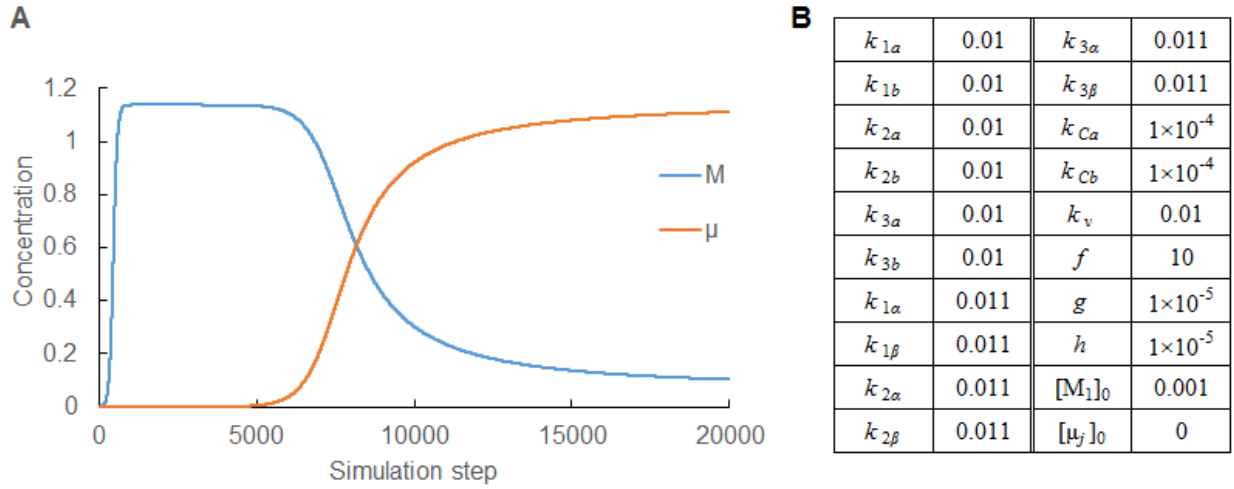

**Figure 14. Succession with deterministic addition of G and H**. **(A)** The dynamics of the total concentration of members for the M and μ cycles, with constant addition of G and H with F. **(B)** The parameters used in the simulation.

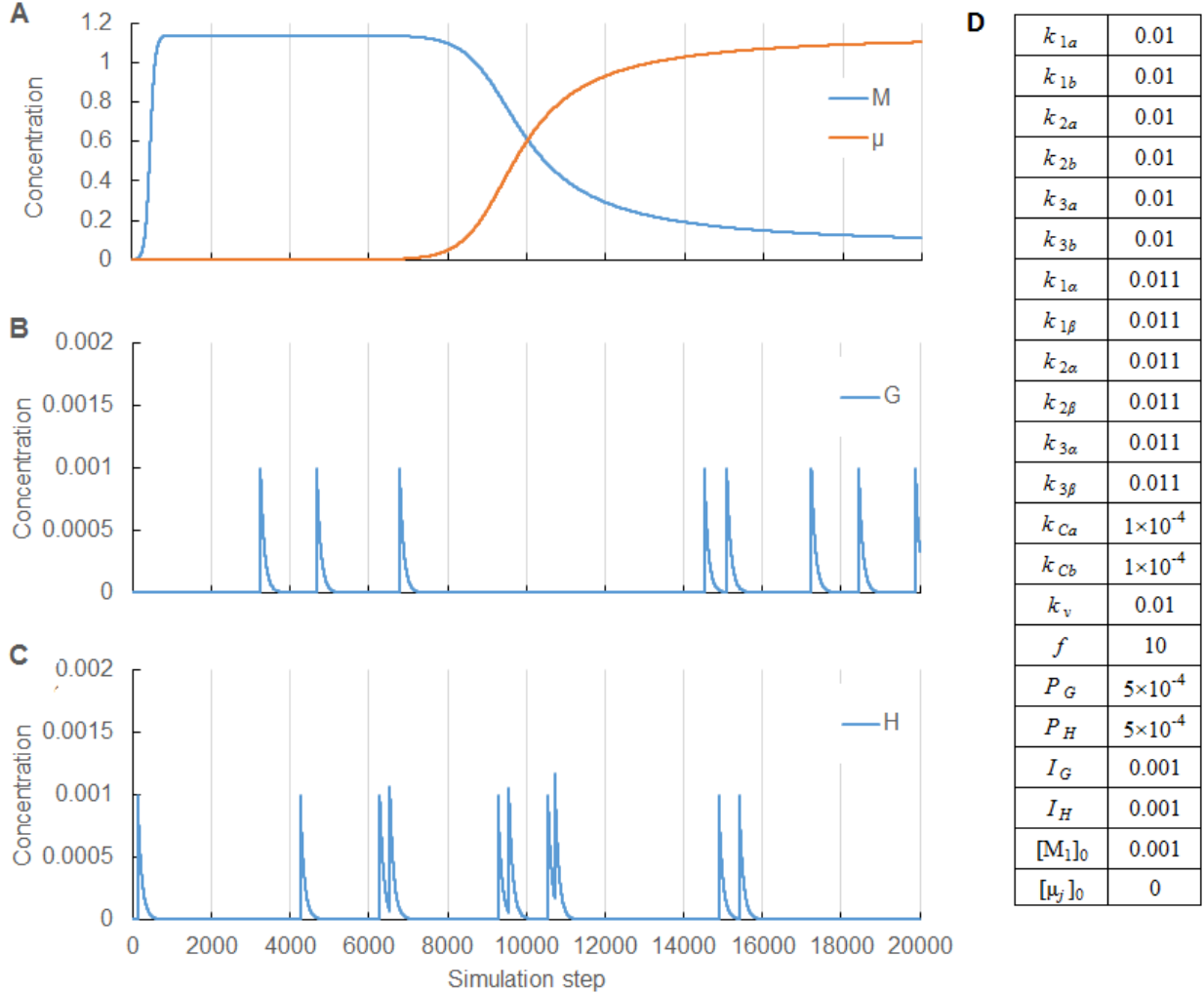

**Figure 15. Succession with stochastic addition of G and H.** (**A**) The dynamics of the total concentration of members for the M and μ cycles. (**B**) The dynamics of G. (**C**) The dynamics of H. (**D**) The parameters used in the simulation.

### 6.2 Occupancy advantage

Even when succession is dependent on stochastic addition of chemicals, as in G and H in the preceding example, the progress may still be deterministic in the sense that, if one waits long enough the μ cycle is sure to take over. Chance events can affect the timing of the event but not the long-term outcome. This contrasts with familiar evolutionary dynamics which are subject to historical contingency: chance events can "push" an evolving lineage along different paths, and these different alternative paths may continue to diverge almost indefinitely. The first place to look for such historical contingency is occupancy advantage, in which an established occupant of an environment has an advantage over potential invaders merely by virtue of having established first (Wright and Vetsigian, 2016). Such a "survival of the first" or "priority effect" phenomenon is important to explore because it establishes cases in which, depending on which cycles are actualized by rare seeding events first, different steady states can be established. Or, to put it another way, an ecosystem can be said to remember prior chemical exposures, at least for a time. In this section, we will show how the mutual inhibition of two 2-0-0 cycles can result in occupancy advantage (Figure 16). The inhibition is mediated by the waste of one cycle reacting with the food of the other cycle to form a compound, X or Y, that cannot be directly utilized by either cycle.

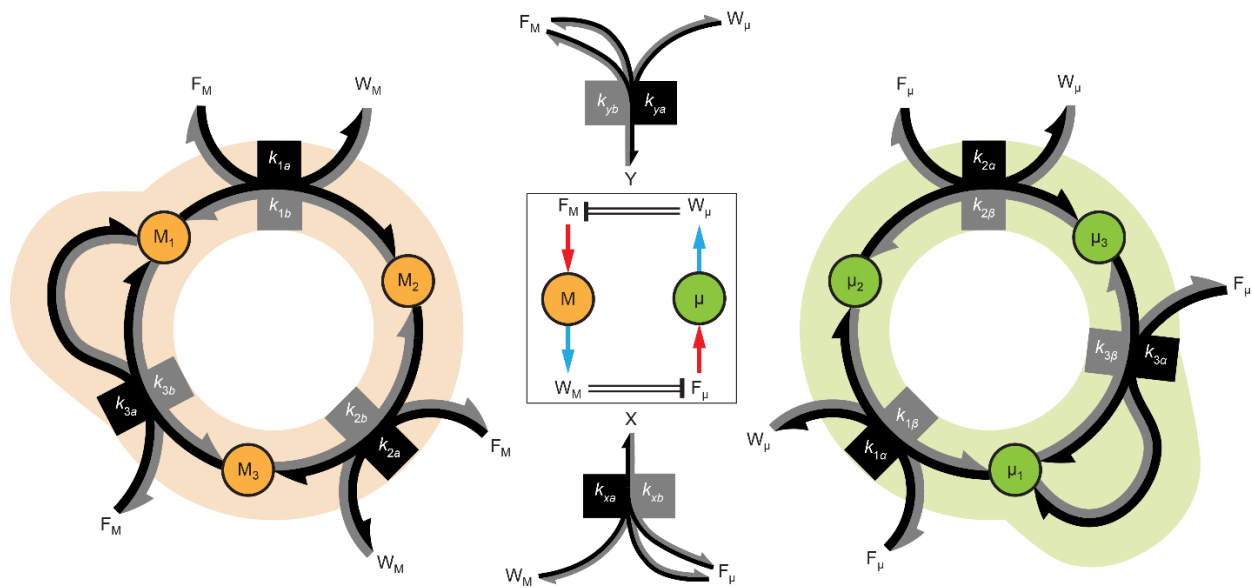

**Figure 16. A possible reaction networks underlying occupancy advantage.** The waste of one cycle reacts with the food of the other cycle to form a compound that cannot be directly utilized by either cycle. Such inhibitory reactions are indicated by the flat-ended double lines in the subpanel, to reflect that the ratio of the food stoichiometric coefficient to that of waste needs to be at least 2:1 for occupancy advantage to occur in this model.

In our simulations we assumed that each cycle has equal rate constants for forward and reverse reactions, but the μ cycle has 10% higher rate constants than the M cycle. Nonetheless, if the M cycle is the first occupant, it can still dominate the reactor and prevent establishment of the μ cycle (Figure 17A); only if the amount of introduced $\mu_1$ is higher than a fixed threshold can the μ cycle take over the reactor (Figure 17B). In other words, the reaction network described in Figure 16 specifies a bistable system, where the non-trivial steady states are either M-dominated or μ-dominated, and the switch between the steady states requires a large-enough trigger. The same holds when the μ cycle is the occupant and the M cycle is the invader (Figure 17C-D), but the threshold amount of $M_1$ for switching the state is much higher.

In some cases when the seeding by members of the M or μ cycles is rare and stochastic, ecosystems can "remember" whether they were first seeded by M or μ. Suppose that a reactor starts to run without any of the members, with occasional influx events bringing in either the $M_1$ or $\mu_1$. Depending on the frequency and magnitude of the influx events, the outcome would be different. When the magnitude of each influx event is lower than the threshold for switching from the M-dominated to the μ-dominated state, the system will nonetheless tend to end up μ-dominated when influx is sufficiently frequent (Figure 18). This is because the system converges to the non-stochastic case of both cycles being seeded equally, except that the μ cycle has a higher carrying capacity and can thus more efficiently utilize the environment and suppress M (Figure 17, $K_M \approx 1.42$, $K_\mu \approx 1.49$). Conversely, if influx is very rare, then the system will converge to whichever state (M-dominated or μ-dominated) triggered by the first influx event (Figure 18). When the magnitude of influx is between the threshold for the M-to-μ switch and that of the μ-to-M switch, the system will end up μ-dominated because the μ cycle can always invade the M-dominated state but not *vice versa*. Finally, given that the magnitude of influx is higher than the threshold for the μ-to-M switch, the system will keep switching between the M-dominated and μ-dominated states because the μ cycle can always invade the M-dominated state and *vice versa*. This shows that the ability of a system to capture information about history (which influx event happened first) depends on the systems dynamics.

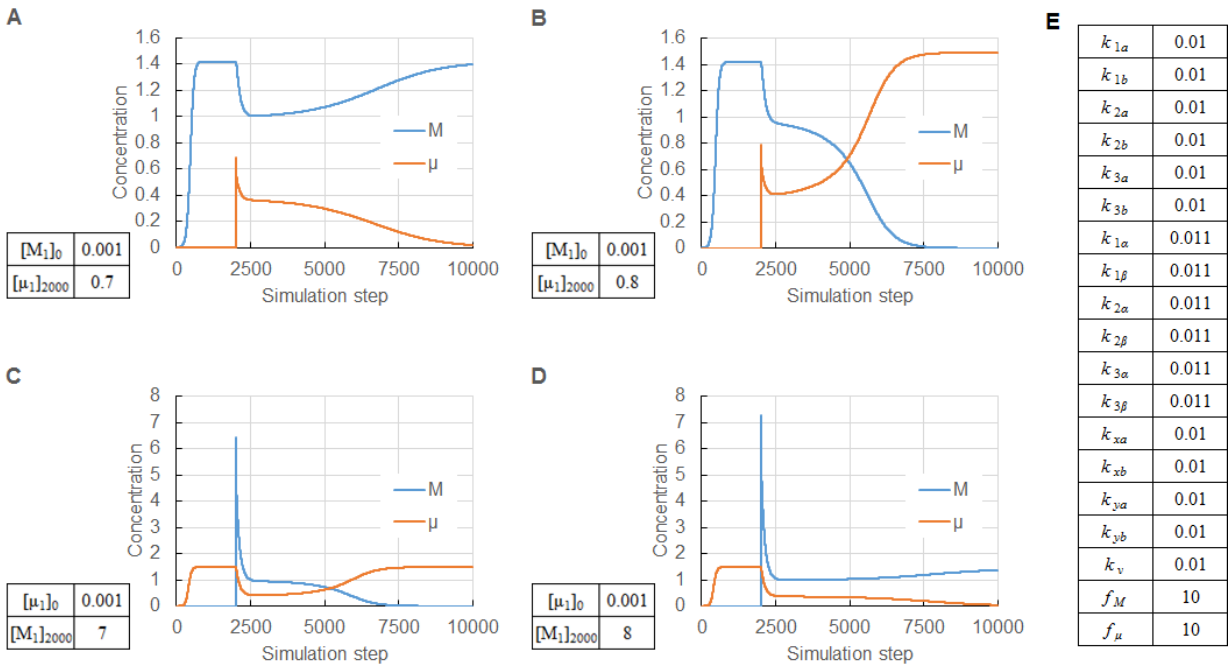

**Figure 17. Occupancy advantage due to mutual inhibition between the autocatalytic cycles not sharing food or waste**. (**A**) (**B**) The dynamics of the total concentration of members for the M and μ cycles, when the M cycle is the occupant and the μ cycle is the invader. If the amount of $\mu_1$ introduced at the 2000th timestep is below the threshold then it cannot invade (**A**), but if it is above the threshold, it can (**B**). (**C**) (**D**) The dynamics of the total concentration of members for the M and μ cycles, when the μ cycle is the occupant and the M cycle is the invader. In this case the concentration of $M_1$ needed to supersede the μ cycle (**C**, **D**) is about ten times higher than that of $\mu_1$ to supersede the M cycle (**A**, **B**). (**E**) Parameters shared by all simulations.

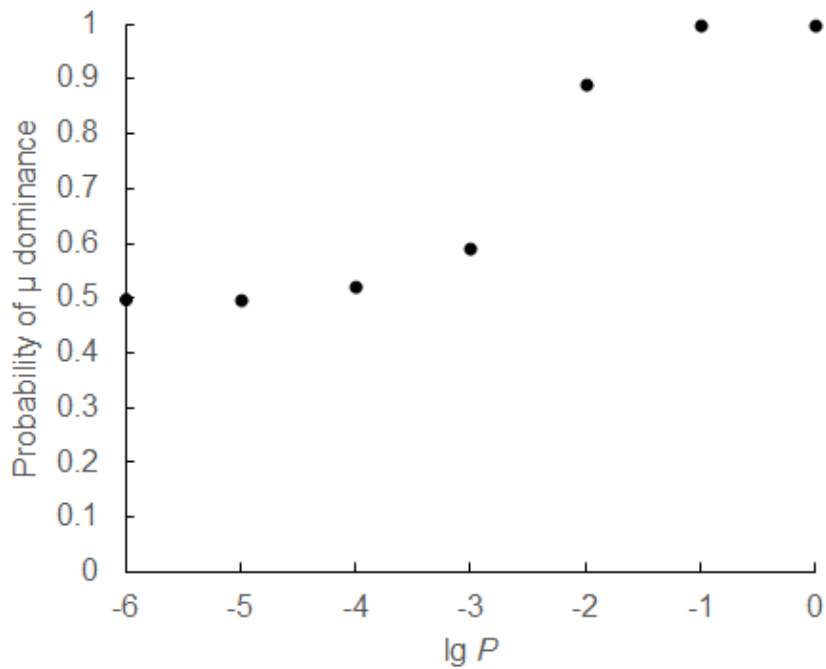

**Figure 18. The frequency of member influx affects the fate of an occupancy-advantage system.** The reactor is initialized without any of the members. The influx of $M_1$ and $\mu_1$ are independent, and the probability of $M_1$ influx per simulation step and that of $\mu_1$ are both $P$. The magnitude of each influx is $10^{-6}$ concentration units. Other parameters are the same as Figure 17E. For each $P$ value, we ran 1000 simulations and calculated the proportion of $\mu$-dominated results (i.e., vertical axis). Note that the horizontal axis shows the common logarithm of $P$.

## DISCUSSION

### Linking ecology, metabolism, and systems biology

It has long been appreciated that there are parallels between chemical reaction systems and ecology (Lloyd, 1967). Likewise, flux balance analysis has been applied to studying ecological communities (Khandelwal et al., 2013). Here we have explored these parallels in more detail, focusing on one particular chemical motif, the autocatalytic cycle.

We show that simple autocatalytic cycles exhibit logistic growth in a flow reactor when the ratio of the food concentration to the dilution rate of the flow is above a threshold and when a reactor is seeded with a small quantity of a member. Furthermore, the form of the logistic equation can be connected back to the parameters describing the flow and the rate constants of the chemical reactions.

The model shows that even when the autocatalytic cycle only consists of reversible reactions whose standard Gibbs energy change are positive (along the autocatalytic direction), the growth to steady-state concentrations can still occur if the flux of food is high enough. This confirms that the flux of food itself can drive a system out of equilibrium. This is why catalysis is important: despite having no effect on the standard Gibbs energy changes, catalysis permits greater flux from external drivers into life-like chemical networks, thereby raising both the intrinsic growth rate and the carrying capacity.

Our explorations suggest that metabolic networks contain multiple autocatalytic cycles. Insofar as these cycles are independent (i.e., non-overlapping), it may be possible to predict their behavior using coarse-graining procedures rooted in ecology. This might include ideas such as trophic levels, food webs, and guilds. This would be useful because even if we know the reactions that can occur in a chemical ecosystem, we often know little about their rate constants, making thorough quantitative analysis difficult.

A complication that would likely arise in coarse-graining procedures is that autocatalytic cycles may share members. Although we have not investigated such overlapping simple autocatalytic cycles in this paper, it might be necessary to find a proper way to describe and analyze them in the future. It is unclear if overlapping autocatalytic cycles should be treated as an integrated larger autocatalytic cycle or two semi-separate cycles, such as when males and females coexist in a single biological species.

**Relevance to the origin of life**

A key problem in the origin-of-life research is whether metabolism or genetics evolved first (Pross, 2004; Vasas et al., 2010; Xavier et al., 2019). Supporting a metabolism-first perspective is the fact that genetic polymers are energetically expensive, unlikely to form by chance, and would seem to require the simultaneous existence of metabolic systems for the continuous formation of their constituent monomers. However, while metabolism-first approaches can readily explain the spontaneous appearance of self-propagating (i.e., autocatalytic) systems, the means by which these systems could evolve adaptively so as to eventually yield genetic systems remain poorly understood. Here we have used the model of an ecosystem of autocatalytic cycles in an energetically driven environment as a starting point for filling this important gap in metabolism-first models.

We have shown that even an ecosystem composed of just two autocatalytic cycles may exhibit complicated long-term dynamics. Such transitions can be due to deterministic features (e.g., competition) and/or stochastic features (e.g., occupancy advantage). The deterministic features generally result in higher efficiency of utilizing the food flowing into the environment to support the survival and growth of the autocatalytic system(s). On the other hand, stochastic features, such as seen in occupancy advantage, illustrate the potential for autocatalytic systems to become trapped in subregions of a multidimensional space where each dimension is the concentration of a chemical, which could readily canalize future changes as additional potential cycles become actualized by rare seeding events. Together these processes suggest that evolution may be viewed as a special combination of different ecological interactions, and that more complex ecosystems could evolve toward local optima by deterministic forces, while nonetheless, being subject to historical constraints.

One way to apply our ecological perspective to investigating the genetics-first *versus* metabolism-first debate is to view self-replicating polymers as the members of autocatalytic cycles, with the branching reaction being a fragmentation or strand separation step and the monomers serving as the food needed for polymer size to increase prior to branching. Even if this is the case, the food monomers would most likely arise from the members of some other autocatalytic cycle(s) (just as some amino acids are derived from members of the reductive citric acid cycle). In that case the polymer cycle would behave like a predator or parasite and the monomer-producing cycle(s) as its prey or host(s). However, such a situation might not be stable in the long run. On one hand, if the predator cycle is not efficient enough, its members would have significantly lower steady-state concentrations, making them more prone to local extinction (loss of all members of the polymer cycle) due to environmental perturbations. On the other hand, if the predator cycle is very efficient in consuming monomers, the total concentration of the predator cycle and the monomer-producing cycle would be largely suppressed, making the entire system more prone to local extinction. The probability of such local extinction would be lower, however, if the polymers provided some reciprocal benefit to the monomer-producing cycles that would raise the local concentrations of both cycles. The most obvious form of such mutualism would be via catalysis: if polymers increased the rate constants of one or more reactions in the monomer-producing cycle upon which they depend, then those polymers together with the monomer-producing cycle would be more extinction-resistant.

To extend this model further and put it to work in explaining the origin of life, at least five additional factors are needed. First, we will need to consider different mechanisms that can generate similar ecological interactions. For example, we showed that mutualisms could be mediated by recycled waste, but it is also possible that catalysis can result in mutualisms. Second, we will need to consider larger and more realistic networks, which would not only include simple autocatalytic cycles but also other motifs such as compound cycles, futile cycles, combinatorial explosions of cross-reactive monomers, and diverse topologies of reaction paths linking these cycles or connecting cycles to the external environment (Hadadi et al., 2016; Michal and Schomburg, 2012). It is therefore important to see if these more realistic chemical ecosystems show ecological and evolution-like dynamics. Third, we will need to consider adsorption-desorption-diffusion processes and add spatial heterogeneity and local disturbances to permit the study of selection. This would allow us to evaluate tradeoffs between different ecosystem "phenotypes"; for example, due to tradeoffs between competitive ability and colonizing ability, selection might prevent divergent systems from converging on one or few locally optimal states. Fourth, we should quantitively explore the models of autocatalytic systems where some reactions are catalyzed by some chemicals in the set, thus generating a more general theory of which the RAF theory (Hordijk et al., 2012; Hordijk and Steel, 2016) is a special case. By understanding the selective environments in which catalysts are favored, we can move towards a better understanding of why life exhibits such a heavy dependence on specific (mainly protein) catalysts. Fifth, it could be instructive to explore polymerization cascades, while allowing that some sequences of polymers act as catalysts of their own production and/or allow that polymerization is intrinsically and weakly template-guided. Taken together, these additional pieces of theory may explain the evolutionary origin of genetic systems as a result of pre-genetic modes of adaptive evolution rooted in chemical ecological interactions, and may help predict the chemical and physical conditions needed to experimentally generate life-like chemical systems.

**FUNDING**

This research was funded by the NASA-NSF CESPOoL (Chemical Ecosystem Selection Paradigm for the Origins of Life, ) Ideas Lab grant (NASA-80NSSC17K0296).

**ACKNOWLEDGEMENTS**

We thank multiple individuals for constructive conversations about evolution in chemical networks, especially, Domenico Bullara, Gheorghe Craciun, Irving Epstein, Chris Kempes, Daniel Segrè, Kalin Vetsigian, Lena Vincent, and John Yin. We also thank all members of the Baum Lab, the CESPOoL consortium (https://cespool.org), and the Santa Fe Institute Evolving Chemical Systems working group.

**MATERIALS AND METHODS**

**1. Simulation algorithm**

For a chemical $\kappa$, its concentration in the reactor is $[\kappa]$. The instantaneous change in $[\kappa]$ may due to three processes: addition, dilution, and reaction.

If $\kappa$ is added constantly through the entrance I (Figure 1C), we have

$$\left.\frac{\mathrm{d}[\kappa]}{\mathrm{d}t}\right|_{\text{addition,i}} = k_v f_\kappa \tag{12}$$

where $k_v$ is the dilution rate and $f_\kappa$ is the concentration of $\kappa$ in the source solution.

If $\kappa$ is added through the port P (Figure 1C) at time $t'$, we have

$$\left.\frac{\mathrm{d}[\kappa]}{\mathrm{d}t}\right|_{\text{addition,iii};t=t'} = I_{\kappa;t=t'} \tag{13}$$

where $I_{\kappa;t=t'}$ is the instantaneous growth rate of $[\kappa]$ only due to the addition of $\kappa$ through the entrance iii at $t'$.

$\kappa$ is constantly diluted and removed from the reactor through the exit O (Figure 1C), so we have

$$\left.\frac{\mathrm{d}[\kappa]}{\mathrm{d}t}\right|_{\text{dilution};t=t'} = -k_v[\kappa]_{t=t'} \tag{14}$$

$\kappa$ is constantly produced and consumed by chemical reactions. For a reversible reaction $R_j$ consisting of the forward reaction $R_{ja}$ and the reverse reaction $R_{jb}$, we have

$$\left.\frac{\mathrm{d}[\kappa]}{\mathrm{d}t}\right|_{R_j;t=t'} = v_{ja;t=t'}\left(-\omega_{ja,\kappa} + \omega_{jb,\kappa}\right) + v_{jb;t=t'}\left(-\omega_{jb,\kappa} + \omega_{ja,\kappa}\right) \tag{15}$$

where $v_{ja}$ and $v_{jb}$ are respectively the reaction rates of $R_{ja}$ and $R_{jb}$ calculated according to the rate law, $\omega_{ja,\kappa}$ and $\omega_{jb,\kappa}$ are respectively the stoichiometric coefficients of $\kappa$ on the reactant and product sides of $R_{ja}$. Specifically, if $\kappa$ is only on the reactant side of $R_{ja}$, then $\omega_{jb,\kappa} = 0$; if $\kappa$ is only on the product side of $R_{ja}$, then $\omega_{ja,\kappa} = 0$; if $\kappa$ is not involved in $R_{ja}$, then $\omega_{ja,\kappa} = \omega_{jb,\kappa} = 0$.

Thus, for the flow reactor where there are $Z$ possible reversible reactions, the instantaneous change rate in $[\kappa]$ at $t'$ is given by

$$\left.\frac{\mathrm{d}[\kappa]}{\mathrm{d}t}\right|_{t=t'} = \left.\frac{\mathrm{d}[\kappa]}{\mathrm{d}t}\right|_{\text{addition,i}} + \left.\frac{\mathrm{d}[\kappa]}{\mathrm{d}t}\right|_{\text{addition,iii};t=t'} + \left.\frac{\mathrm{d}[\kappa]}{\mathrm{d}t}\right|_{\text{dilution};t=t'} + \sum_{j=1}^{Z}\left.\frac{\mathrm{d}[\kappa]}{\mathrm{d}t}\right|_{R_j;t=t'} \tag{16}$$

Equation (16) needs to be applied to every chemical involved in at least one of the $Z$ reversible reactions.

To run numerical simulations, the instantaneous change in $[\kappa]$ from $t_i$ to $t_{i+1}$ is approximated by

$$\Delta[\kappa]_{t_i \rightarrow t_{i+1}} = \Delta t \cdot \left.\frac{\mathrm{d}[\kappa]}{\mathrm{d}t}\right|_{t=t_i} \tag{17}$$

where the time step size $\Delta t = t_{i+1} - t_i$. For simplification, in all the simulations reported in this paper, $\Delta t = 1$.

The simplification that $\Delta t = 1$ would cause errors when $\mathrm{d}[\kappa]/\mathrm{d}t$ is too large. When such an error was detected, we decreased the rate constants and dilution rates proportionally to circumvent the error.

The Python code that we used to perform the simulations is in Supplemental Materials: Python Code.

**SUPPLEMENTAL MATERIALS**

## 1. The dynamics of the 0-0-0 cycle can be approximated by a logistic model

The differential equations describing the dynamics of the concentrations of M and F in the reactor are

$$\begin{cases} \dfrac{\mathrm{d[M]}}{\mathrm{d}t} = -k_v[M] + k_a[M][F] - k_b[M]^2 \\ \dfrac{\mathrm{d[F]}}{\mathrm{d}t} = k_v f - k_v[F] - k_a[M][F] + k_b[M]^2 \end{cases} \tag{S1}$$

If we define $x = [M] + [F]$, we can get

$$x(t) = f - c_0 e^{-k_v t} \tag{S2}$$

Where $c_0$ is a constant and its value depends on the initial conditions of the reactor. Now we assume that M is added into the reactor with the concentration $[M]_0 \ll f$, then we have

$$x(t = 0) = f - c_0 e^{-k_v \cdot 0} = f - c_0 = f + [M]_0 \approx f \tag{S3}$$

Thus, $c_0 \approx 0$, which also means that $[M] + [F] \approx f$. With this approximation, we can rewrite the differential equation describing the dynamics of the member as

$$\dfrac{\mathrm{d[M]}}{\mathrm{d}t} = -k_v[M] + k_a[M](f - [M]) - k_b[M]^2 = (k_a f - k_v)[M] - (k_a + k_b)[M]^2 \tag{S4}$$

By defining $r_M = k_a f - k_v$, and $K_M = (k_a f - k_v) / (k_a + k_b)$, this equation can be written in a logistic form:

$$\dfrac{\mathrm{d[M]}}{\mathrm{d}t} = r_M[M]\left(\dfrac{K_M - [M]}{K_M}\right) \tag{3}$$

## 2. The intrinsic growth rate, carrying capacity, and potential dispersal of the 0-0-0 cycle are intrinsically associated and are affected by catalysis

First, although the intrinsic growth rate and the carrying capacity are usually treated as independent constants by ecologists, our analysis shows that $r_M$ and $K_M$ for the 0-0-0 cycle are not independent. According to Equation (4), we can infer that any factors increasing $r_M$ will also increase $K_M$, while the factors increasing $k_b$ will decrease $K_M$ but have no effects on $r_M$. In other words, it is almost impossible to modify $r_M$ while keeping $K_M$ unchanged (unless the effect of changed $r_M$ happens to be exactly compensated by a change of $k_b$), but it is not that hard to modify $K_M$ while keeping $r_M$ unchanged.

Second, although catalysis, which is to say a symmetric reduction in the activation energies of the forward and reverse reactions, does not change the $k_a$-to-$k_b$ ratio (according to the definition of catalysis), it *will* still affect $r_M$ and $K_M$. If $k_b = p \cdot k_a$ then, according to Equation (4), we get:

$$K_M = \frac{f}{1+p} - \frac{k_v}{1+p} \cdot \frac{1}{k_a} \tag{S5}$$

Supposing that a catalyst can decrease the activation energies of the 0-0-0 cycle, such that both $k_a$ and $k_b$ are increased but $p$ does not change. Then, according to Equations (4) and (S5), we can conclude that adding such a catalyst continuously through the entrance I (Figure 1C) into the reactor will increase both $r_M$ and $K_M$.

Also, whereas $r_M$ does not have an upper limit when catalysts are added, $K_M$ has an upper limit of $f \,/\, (1 + p)$, or $f \,/\, 2$ in the case of symmetric rate constants (i.e., reactions that are not thermodynamically driven). These conclusions also suggest that the more efficient a catalyst, the faster the system's growth and the higher its steady state concentration ($K$). In addition, if the rate constants are mapped to activation energies by the Arrhenius equation (Supplemental Materials, Section 3), which specifies activation energies ($E_a$ and $E_b$ for the forward and reverse reactions, respectively) and pre-exponential factors ($A_a$ and $A_b$ for the forward and reverse reactions, respectively), it can be shown that $r_M$ is more sensitive to a change in the efficiency of catalysis when catalysts are already quite efficient, such that $E_a$ is small (Figure S1A), whereas $K_M$ is more sensitive to a change in the efficiency of catalysis when $E_a$ is large (Figure S1B-C).

### 3. The relationship between the growth dynamics and activation energies of a 0-0-0 cycle

For a 0-0-0 cycle, the relationship between the rate constants and growth dynamics is given by

$$\frac{d[M]}{dt} = r_M[M] \left( \frac{K_M - [M]}{K_M} \right) \tag{3}$$

$$\begin{cases} r_M = k_a f - k_v \\ K_M = \dfrac{k_a f - k_v}{k_a + k_b} \end{cases} \tag{4}$$

If we use Arrhenius equation to describe the relationship between the rate constants, such that

$$\begin{cases} k_a = A_a e^{\frac{-E_a}{RT}} \\ k_b = A_b e^{\frac{-E_b}{RT}} \end{cases} \tag{S6}$$

where $k_a$ and $k_b$ are rate constants, $A_a$ and $A_b$ are pre-exponential factors, $E_a$ and $E_b$ are activation energies for the forward and reverse reactions, $R$ is the gas constant, and $T$ is thermodynamic temperature. Thus, Equation (4) can be rewritten as

$$\begin{cases} r_M = A_a e^{\frac{-E_a}{RT}} \cdot f - k_v \\ K_M = \dfrac{A_a e^{\frac{-E_a}{RT}} \cdot f - k_v}{A_a e^{\frac{-E_a}{RT}} + A_b e^{\frac{-E_b}{RT}}} \end{cases} \qquad \text{(S7)}$$

If $E_a$, $E_b$, $r_M$, $K_M$ are variables and all other parameters are constants, we can plot the $r_M$ - $E_a$ curve and $K_M$ - $E_a$ - $E_b$ surface (Figure S1A-B). Specifically, in Figure S1B, if we cut the $K_M$ - $E_a$ - $E_b$ surface by a plane defined by

$$E_a - E_b = \Delta_r G^o \qquad \text{(S8)}$$

where $\Delta_r G^o$ is the standard Gibbs energy change of the forward reaction, then the intersection of the plane and the $K_M$ - $E_a$ - $E_b$ surface represents how $K_M$ changes with different efficiencies of catalysis. Thus, we can plot the projection of this intersection onto the $E_a$ - $K_M$ plane, and the resulting curve can be used to assess the relationship between $K_M$ and the efficiency of catalysis (Figure S1C). $r_M$ is more sensitive to the change in the efficiency of catalysis when $E_a$ is small (Figure S1A), likely corresponding to the transition from moderately efficient catalysis to highly efficient catalysis. On the other hand, $K_M$ is more sensitive to the change in the efficiency of catalysis when $E_a$ is large, likely corresponding to the transition from no catalysis to the presence of catalysis (Figure S1B-C).

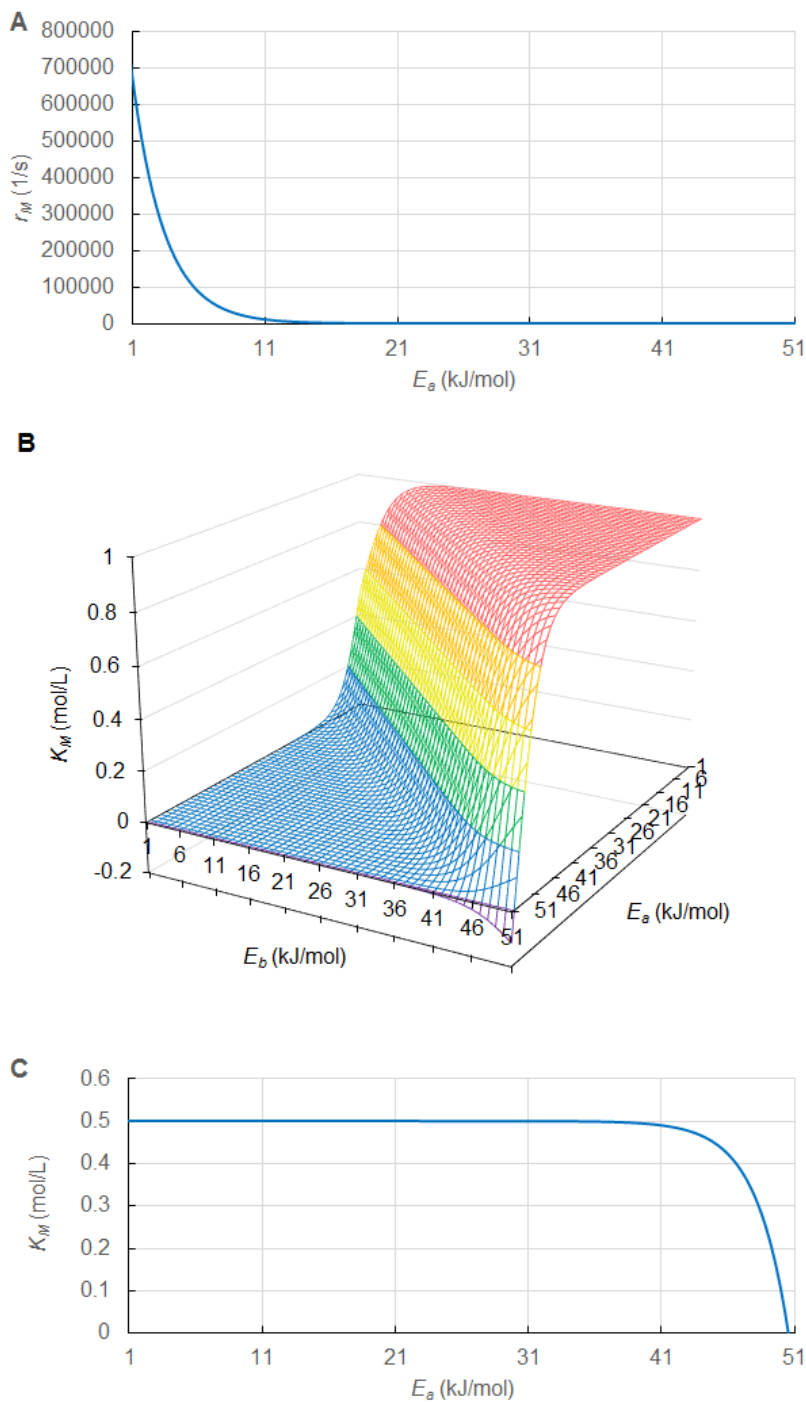

**Figure S1. The relationship between the efficiency of catalysis and the growth dynamics of a 0-0-0 cycle.** The parameters used to generate the graphs are: pre-exponential factors $A_a = A_b = 10^6$ mol$^{-1}$·L·s$^{-1}$, $k_v = 0.001$ s$^{-1}$, $f = 1$ mol·L$^{-1}$, the gas constant $R = 8.314$ J·mol$^{-1}$·K$^{-1}$, temperature $T = 293.15$ K. Note that the efficiency of catalysis is negatively correlated to the activation energy. **(A)** The relationship between the $E_a$ (the activation energy of the forward reaction) and $r_M$. **(B)** The relationship between $E_a$, $E_b$ (the activation energy of the reverse reaction) and $K_M$. **(C)** The relationship between $E_a$ and $K_M$, given that $\Delta_r G^o = E_a - E_b = 0$ kJ·mol$^{-1}$.

## 4. The maximum global dispersal of the 0-0-0 cycle will be achieved when $k_v / k_a = f / 2$

We know that at the steady state, $[M] = K_M$. Then according to Equation (4), we get:

$$k_v[M] = \frac{k_a f k_v - k_v^2}{k_a + k_b} \tag{S9}$$

According to the properties of a quadratic function, the maximum $k_v[M]$ is achieved when

$$k_v = -\frac{k_a f}{2 \cdot (-1)} = \frac{k_a f}{2} \tag{S10}$$

which is equivalent to $k_v / k_a = f / 2$.

## 5. The dynamics of the 1-0-0 cycle can be approximated by a logistic model

The differential equations describing the dynamics of the concentrations of $M_1$, $M_2$, F, and W in the reactor are

$$\begin{cases} \dfrac{d[M_1]}{dt} = -k_v[M_1] - k_{1a}[M_1][F] + k_{1b}[M_2][W] + 2k_{2a}[M_2][F] - 2k_{2b}[M_1]^2 \\[2mm] \dfrac{d[M_2]}{dt} = -k_v[M_2] + k_{1a}[M_1][F] - k_{1b}[M_2][W] - k_{2a}[M_2][F] + k_{2b}[M_1]^2 \\[2mm] \dfrac{d[W]}{dt} = -k_v[W] + k_{1a}[M_1][F] + k_{1b}[M_2][W] \\[2mm] \dfrac{d[F]}{dt} = k_v f - k_v[F] - k_{1a}[M_1][F] + k_{1b}[M_2][W] - k_{2a}[M_2][F] + k_{2b}[M_1]^2 \end{cases} \tag{S11}$$

Similar to the 0-0-0 cycle, it can be shown that $[M_1] + [M_2] + [W] + [F] \approx f$. Then by summing the first three rows in Equation (S11) and making the substitutions:

$$\begin{cases} S = [M_1] + [M_2] + [W] \\ [M_1] = \theta_1 S \\ [M_2] = \theta_2 S \\ [W] = (1 - \theta_1 - \theta_2)S \end{cases} \tag{5}$$

we can get

$$\frac{dS}{dt} = (k_{1a}\theta_1 f + k_{2a}\theta_2 f - k_v)S - (k_{1a}\theta_1 + k_{2a}\theta_2 + k_{1b}\theta_2(1 - \theta_1 - \theta_2) + k_{2b}\theta_1^2)S^2 \tag{S12}$$

This equation can be written in the logistic form

$$\frac{dS}{dt} = r_S S \left( \frac{K_S - S}{K_S} \right) \tag{6}$$

by defining:

$$\begin{cases} r_S = k_{1a}\theta_1 f + k_{2a}\theta_2 f - k_v \\ K_S = \dfrac{k_{1a}\theta_1 f + k_{2a}\theta_2 f - k_v}{k_{1a}\theta_1 + k_{2a}\theta_2 + k_{1b}\theta_2(1 - \theta_1 - \theta_2) + k_{2b}\theta_1^2} \end{cases} \qquad (7)$$

Similarly, it can be shown that the dynamics of [M$_1$], [M$_2$], and [W] can all be written in logistic forms.

## 6. Initiating members affects the growth dynamics

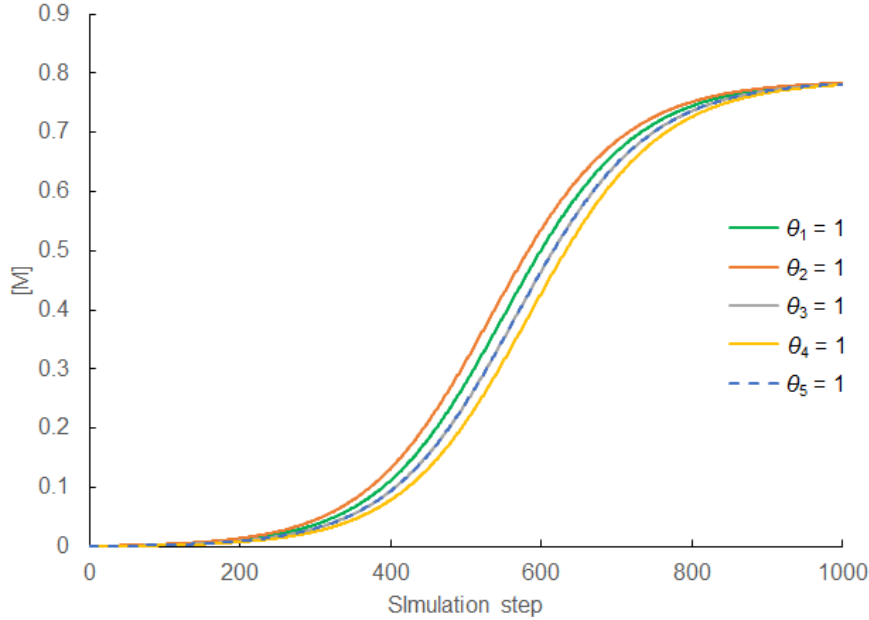

**Figure S2. Initiating members of a 1-1-2 cycle affects the growth dynamics**. The dynamics of the total concentration of members were generated by setting all rate constants to 0.01, $k_v = 0.01$, $f = 10$, and initiating the reactions with different members. For example, $\theta_1 = 1$ corresponds to [M$_1$]$_0$ = 0.001 and [M$_2$]$_0$ = [M$_3$]$_0$ = [M$_4$]$_0$ = [M$_5$]$_0$ = [W]$_0$ = 0.

## 7. The relationship between $r_M$, $K_M$, $k_v K_M$ and the combination of $f$ and $k_v$

For each of the 2-0-0, 2-0-1, and 2-1-1 cycles, we know that approximately, the cycle can grow and sustain only when $f / k_v$ is higher than a threshold value $\eta$ (Figure 5). By anticlockwise rotating the coordinate system where $k_v$ and $f$ are respectively the horizontal and vertical axes, such that the horizontal axis overlaps the line defined by $f = \eta k_v$, it is easy to show that the vertical coordinate of this rotated coordinate system is given by $(f - \eta k_v)/\sqrt{1 + \eta^2}$. We may define that

$$\varphi = \frac{f - \eta k_v}{\sqrt{1 + \eta^2}} \qquad (S13)$$

Then we can plot $r_M$, $K_M$, and $k_v K_M$ against $\varphi$ (Figure S3).

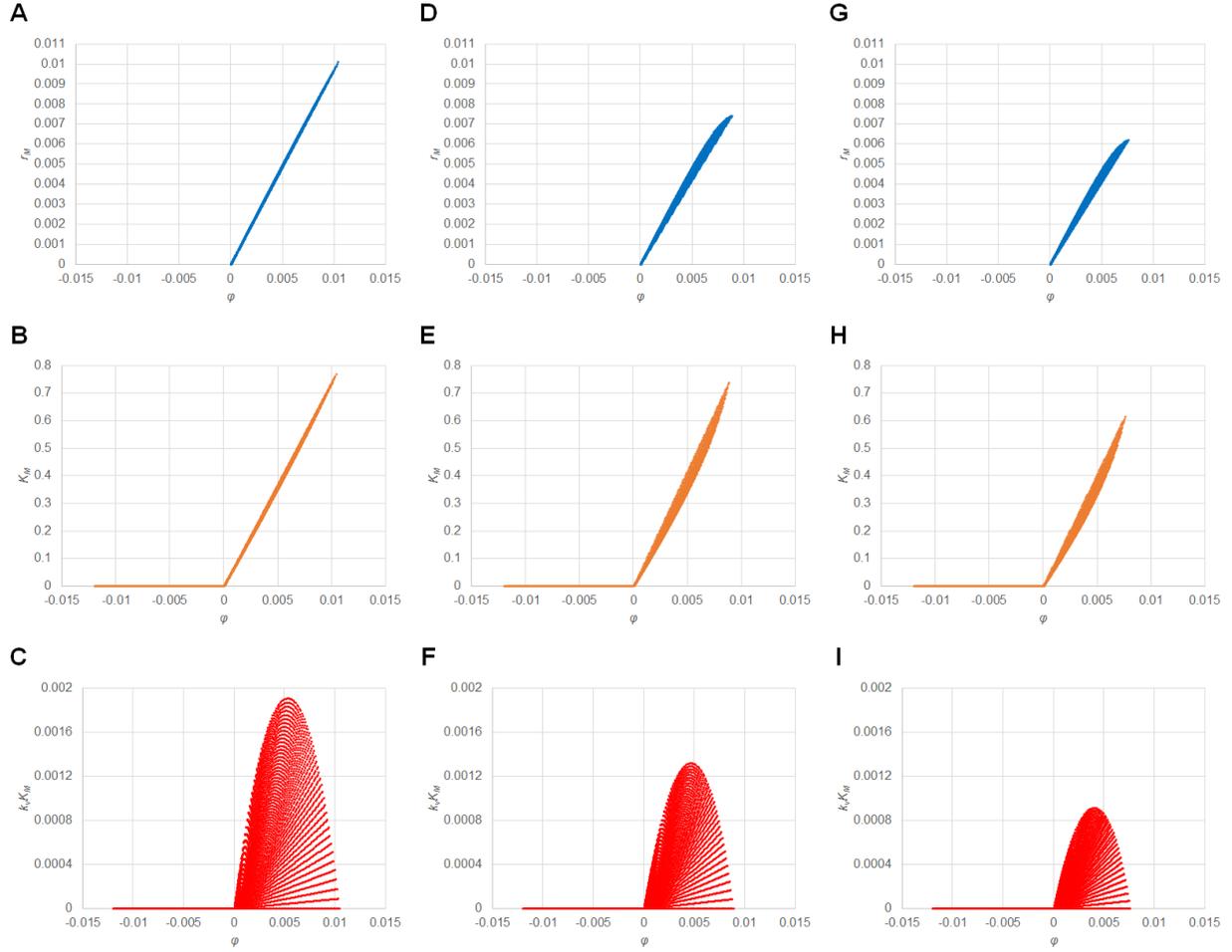

**Figure S3. The relationship between $r_M$, $K_M$, $k_v K_M$ and $\varphi$.** For all data points used to generate Figure 5, the $\varphi$ values were calculated, and then $r_M$, $K_M$, and $k_v K_M$ were plotted against $\varphi$. **(A)** - **(C)** 2-0-0 cycle; $\eta = 384.6$. **(D)** - **(F)** 2-0-1 cycle; $\eta = 452.9$. **(G)** - **(I)** 2-1-1 cycle; $\eta = 528.2$.

## 8. Possible equilibria of the competition between two 0-0-0 cycles

The differential equations describing the dynamics of the concentrations of M, μ, and F in the reactor are

$$\begin{cases} \dfrac{\mathrm{d}[\mathrm{M}]}{\mathrm{d}t} = -k_v[\mathrm{M}] + k_a[\mathrm{M}][\mathrm{F}] - k_b[\mathrm{M}]^2 \\[2mm] \dfrac{\mathrm{d}[\mu]}{\mathrm{d}t} = -k_v[\mu] + k_\alpha[\mu][\mathrm{F}] - k_\beta[\mu]^2 \\[2mm] \dfrac{\mathrm{d}[\mathrm{F}]}{\mathrm{d}t} = k_v f - k_v[\mathrm{F}] - k_a[\mathrm{M}][\mathrm{F}] + k_b[\mathrm{M}]^2 - k_\alpha[\mu][\mathrm{F}] + k_\beta[\mu]^2 \end{cases} \quad \text{(S14)}$$

By applying the approximation $[M] + [\mu] + [F] \approx f$, we can get

$$\begin{cases} \dfrac{\mathrm{d}[M]}{\mathrm{d}t} = (k_a f - k_v)[M] - (k_a + k_b)[M]^2 - k_a[M][\mu] \\ \dfrac{\mathrm{d}[\mu]}{\mathrm{d}t} = (k_\alpha f - k_v)[\mu] - (k_\alpha + k_\beta)[\mu]^2 - k_\alpha[M][\mu] \end{cases} \quad \text{(S15)}$$

Let both rows of Equation (S15) equal to zero, and we will get a system of quadratic equations

$$\begin{cases} (k_a f - k_v)[M] - (k_a + k_b)[M]^2 - k_a[M][\mu] = 0 \\ (k_\alpha f - k_v)[\mu] - (k_\alpha + k_\beta)[\mu]^2 - k_\alpha[M][\mu] = 0 \end{cases} \quad \text{(S16)}$$

Equation (S16) specifies the relationship between $[M]$ and $[\mu]$ when the system reaches a steady state. By solving Equation (S16), we will get four pairs of solutions.

The first pair is

$$\begin{cases} [M] = 0 \\ [\mu] = 0 \end{cases} \quad \text{(S17)}$$

which means none of the competitors survives, and therefore this pair was not considered in the main text.

The second pair is

$$\begin{cases} [M] = \dfrac{k_a k_\beta f - k_v(k_\alpha + k_\beta - k_a)}{k_a k_\beta + k_\alpha k_b + k_b k_\beta} \\ [\mu] = \dfrac{k_\alpha k_b f - k_v(k_a + k_b - k_\alpha)}{k_a k_\beta + k_\alpha k_b + k_b k_\beta} \end{cases} \quad \text{(8)}$$

The third pair is

$$\begin{cases} [M] = \dfrac{k_a f - k_v}{k_a + k_b} \\ [\mu] = 0 \end{cases} \quad \text{(9)}$$

The fourth pair is

$$\begin{cases} [M] = 0 \\ [\mu] = \dfrac{k_\alpha f - k_v}{k_\alpha + k_\beta} \end{cases} \quad \text{(10)}$$

```
#====Python Code========

'''

README:

Copy this script to a text editor, and replace every ~ with either a tab or 4
spaces. Then save the file as PythonCode.py.

This is the script used to generate the dynamics of occupancy advantage. It
has all modules necessary for generating other dynamics/curves/data shown in
the paper.

For getting a single dynamics/curve, you just need to change/add/delete the
relevant parameters to do other simulations.

For running multiple simulations, you may need to loop through different
parameter settings.

Usage:

python PythonCode.py >
RateLawTemplate_flow_reactor_occupancy_advantage_output.txt

'''

import numpy as np

import sys

def
Batch_calc_concentration_profile_after_input_reaction_removal(molecule_name_l
ist, reaction_settings, current_concentration_profile, solubility_settings,
dilution_rate, input_concentration_profile, time_interval):

~concentration_profile_next_step = {}

~#phase_1: input

~dConcentration_over_dt_due_to_input_profile =
batch_calc_dConcentration_over_dt_due_to_input(input_concentration_profile,
dilution_rate)

~#phase_2: reaction
```

```
~reaction_rate_profile = batch_calc_reaction_rates(reaction_settings,
current_concentration_profile)

~dConcentration_over_dt_due_to_reaction_profile =
batch_calc_dConcentration_over_dt_due_to_reaction(molecule_name_list,
reaction_settings, reaction_rate_profile)

~#phase_3: removal/exit

~dConcentration_over_dt_due_to_removal_profile =
batch_calc_dConcentration_over_dt_due_to_removal(current_concentration_profil
e, dilution_rate)

~#update

~concentration_profile_after_removal =
update_concentration_profile(current_concentration_profile,
dConcentration_over_dt_due_to_input_profile,
dConcentration_over_dt_due_to_reaction_profile,
dConcentration_over_dt_due_to_removal_profile, time_interval)

~#precipitation

~for molecule_name in concentration_profile_after_removal:

~~solubility = solubility_settings[molecule_name]

~~calc_conc = concentration_profile_after_removal[molecule_name]

~~corrected_concentration = min(solubility, calc_conc)

~~concentration_profile_next_step[molecule_name] = corrected_concentration

~return(concentration_profile_next_step)

def
batch_calc_dConcentration_over_dt_due_to_input(input_concentration_profile,
dilution_rate):

~dC_over_dt_due_to_input_profile = {}

~for molecule_name in input_concentration_profile:

~~dC_over_dt_due_to_input_profile[molecule_name] =
dilution_rate*input_concentration_profile[molecule_name]

~return(dC_over_dt_due_to_input_profile)
```

```python
def batch_calc_reaction_rates(reaction_settings,
current_concentration_profile):
~reaction_rate_profile = {}
~for reaction in reaction_settings:
~~reaction_setting = reaction_settings[reaction]
~~reaction_rate_constant = reaction_setting[0]
~~reactant_name_list = reaction_setting[1]
~~reactant_molar_list = reaction_setting[2]

~~reaction_rate = reaction_rate_constant
~~for i in range(len(reactant_name_list)):
~~~if reactant_name_list[i] == "H2O":
~~~~reaction_rate = reaction_rate
~~~else:
~~~~reactant_concentration =
current_concentration_profile[reactant_name_list[i]]
~~~~reactant_molar = reactant_molar_list[i]
~~~~reaction_rate = reaction_rate*(reactant_concentration**reactant_molar)
~~reaction_rate_profile[reaction] = reaction_rate

~return(reaction_rate_profile)

def batch_calc_dConcentration_over_dt_due_to_reaction(molecule_name_list,
reaction_settings, reaction_rate_profile):
~dC_over_dt_due_to_reaction_profile = {}
~for focal_molecule_name in molecule_name_list:
~~dC_over_dt_due_to_reaction_profile[focal_molecule_name] =
calc_dConcentration_over_dt_due_to_reaction(focal_molecule_name,
reaction_settings, reaction_rate_profile)
~return(dC_over_dt_due_to_reaction_profile)
```

```python
def calc_dConcentration_over_dt_due_to_reaction(focal_molecule_name,
reaction_settings, reaction_rate_profile):
~dC_over_dt_due_to_reaction = 0
~for reaction in reaction_settings:
~~reactant_name_list = reaction_settings[reaction][1]
~~reactant_molar_list = reaction_settings[reaction][2]
~~product_name_list = reaction_settings[reaction][3]
~~product_molar_list = reaction_settings[reaction][4]
~~reaction_rate = reaction_rate_profile[reaction]

~~for i in range(len(reactant_name_list)):
~~~if focal_molecule_name == reactant_name_list[i]:
~~~~dC_over_dt_due_to_reaction = dC_over_dt_due_to_reaction -
reaction_rate*reactant_molar_list[i]
~~for j in range(len(product_name_list)):
~~~if focal_molecule_name == product_name_list[j]:
~~~~dC_over_dt_due_to_reaction = dC_over_dt_due_to_reaction +
reaction_rate*product_molar_list[j]

~if focal_molecule_name == "H2O":
~~dC_over_dt_due_to_reaction = 0

~return(dC_over_dt_due_to_reaction)

def
batch_calc_dConcentration_over_dt_due_to_removal(current_concentration_profil
e, dilution_rate):
~dC_over_dt_due_to_removal_profile = {}
~for molecule_name in current_concentration_profile:
~~dC_over_dt_due_to_removal_profile[molecule_name] = -
dilution_rate*current_concentration_profile[molecule_name]
```

```
~return(dC_over_dt_due_to_removal_profile)

def update_concentration_profile(current_concentration_profile,
dConcentration_over_dt_due_to_input_profile,
dConcentration_over_dt_due_to_reaction_profile,
dConcentration_over_dt_due_to_removal_profile, time_interval):

~all_involved_molecule_name_list = []

~for molecule_name in current_concentration_profile:

~~if molecule_name not in all_involved_molecule_name_list:

~~~all_involved_molecule_name_list.append(molecule_name)

~for molecule_name in dConcentration_over_dt_due_to_input_profile:

~~if molecule_name not in all_involved_molecule_name_list:

~~~all_involved_molecule_name_list.append(molecule_name)

~for molecule_name in dConcentration_over_dt_due_to_reaction_profile:

~~if molecule_name not in all_involved_molecule_name_list:

~~~all_involved_molecule_name_list.append(molecule_name)

~for molecule_name in dConcentration_over_dt_due_to_removal_profile:

~~if molecule_name not in all_involved_molecule_name_list:

~~~all_involved_molecule_name_list.append(molecule_name)

~updated_concentration_profile = {}

~for molecule_name in all_involved_molecule_name_list:

~~conc = get_record(molecule_name, current_concentration_profile)

~~d_input = get_record(molecule_name,
dConcentration_over_dt_due_to_input_profile)

~~d_reaction = get_record(molecule_name,
dConcentration_over_dt_due_to_reaction_profile)

~~d_removal = get_record(molecule_name,
dConcentration_over_dt_due_to_removal_profile)

~~

~~tmp_conc = conc + (d_input + d_reaction + d_removal)*time_interval
```

```python
~~if tmp_conc<0:
~~~print("##!##Error: negative concentration!!!!!")
~~~sys.exit()
~~else:
~~~updated_concentration_profile[molecule_name] = tmp_conc

~return(updated_concentration_profile)

def get_record(key, target_dict):
~if key in target_dict:
~~record = target_dict[key]
~else:
~~record = 0
~return(record)

def get_present_molecule_name_list(reaction_settings):
~molecule_name_list = []
~for reaction in reaction_settings:
~~reactant_name_list = reaction_settings[reaction][1]
~~product_name_list = reaction_settings[reaction][3]
~~for x in reactant_name_list:
~~~if x not in molecule_name_list:
~~~~molecule_name_list.append(x)
~~for y in product_name_list:
~~~if y not in molecule_name_list:
~~~~molecule_name_list.append(y)
~return(molecule_name_list)
```

```python
def simu_network_switch(steps, initial_conc_settings, reaction_settings,
solubility_settings, dilution_rate, input_concentration_profile,
time_interval):

~molecule_name_list = get_present_molecule_name_list(reaction_settings)

~# initialization

~current_concentration_profile = {}

~for name in initial_conc_settings:

~~if name in molecule_name_list:

~~~current_concentration_profile[name] = initial_conc_settings[name]

~# simulation

~# print concentrations

~print("###Simu_step", end='\t')

~print(*molecule_name_list, sep='\t')

~print("0", end='\t')

~for x in molecule_name_list:

~~print(str(current_concentration_profile[x]), end='\t')

~print()

~for simu in range(steps):

~~# injection module

~~if simu == 2000:# if you want stochastic injection, you may use the
numpy.random.choice() method to determine whether the injection happens in
this step

~~~seed = "Miu1"

~~~invaded_amount = 0.8

~~~current_concentration_profile[seed] = current_concentration_profile[seed]
+ invaded_amount

~~next_concentration_profile =
Batch_calc_concentration_profile_after_input_reaction_removal(molecule_name_l
ist, reaction_settings, current_concentration_profile, solubility_settings,
dilution_rate, input_concentration_profile, time_interval)
```

```
~~current_concentration_profile = next_concentration_profile

~~print(str(simu+1), end='\t')
~~for x in molecule_name_list:
~~~print(str(current_concentration_profile[x]), end='\t')
~~print()

~return(current_concentration_profile)

#--------

k_a = 0.01
k_b = 0.01

k_alpha = 0.011
k_beta = 0.011

k_xa = 0.01
k_xb = 0.01

k_ya = 0.01
k_yb = 0.01

reaction_settings = {
# reaction_name:[rate_constant, [reactant_names], [reactant_coefficients],
[product_names], [product_coefficients]]
"R_M1a":[k_a, ["M1","F_M"], [1,1], ["M2","W_M"], [1,1]],
```

```
"R_M1b":[k_b, ["M2","W_M"], [1,1], ["M1","F_M"], [1,1]],
"R_M2a":[k_a, ["M2","F_M"], [1,1], ["M3","W_M"], [1,1]],
"R_M2b":[k_b, ["M3","W_M"], [1,1], ["M2","F_M"], [1,1]],
"R_M3a":[k_a, ["M3","F_M"], [1,1], ["M1"], [2]],
"R_M3b":[k_b, ["M1"], [2], ["M3","F_M"], [1,1]],

"R_Xa":[k_xa, ["W_M","F_Miu"], [1,2], ["X"], [1]],
"R_Xb":[k_xb, ["X"], [1], ["W_M","F_Miu"], [1,2]],

"R_Miu1a":[k_alpha, ["Miu1","F_Miu"], [1,1], ["Miu2","W_Miu"], [1,1]],
"R_Miu1b":[k_beta, ["Miu2","W_Miu"], [1,1], ["Miu1","F_Miu"], [1,1]],
"R_Miu2a":[k_alpha, ["Miu2","F_Miu"], [1,1], ["Miu3","W_Miu"], [1,1]],
"R_Miu2b":[k_beta, ["Miu3","W_Miu"], [1,1], ["Miu2","F_Miu"], [1,1]],
"R_Miu3a":[k_alpha, ["Miu3","F_Miu"], [1,1], ["Miu1"], [2]],
"R_Miu3b":[k_beta, ["Miu1"], [2], ["Miu3","F_Miu"], [1,1]],

"R_Ya":[k_ya, ["W_Miu","F_M"], [1,2], ["Y"], [1]],
"R_Yb":[k_yb, ["Y"], [1], ["W_Miu","F_M"], [1,2]]
}

# all highly soluble
solubility_settings = {
"M1": 1000000,
"M2": 1000000,
"M3": 1000000,

"Miu1": 1000000,
"Miu2": 1000000,
"Miu3": 1000000,
```

```python
    "F_M": 1000000,

    "W_M": 1000000,

    "F_Miu": 1000000,

    "W_Miu": 1000000,

    "X": 1000000,

    "Y": 1000000
    }

f_M = 10

f_Miu = 10

# starting conditions
initial_conc_settings = {
    "M1": 0.001,

    "M2": 0,

    "M3": 0,

    "Miu1": 0,

    "Miu2": 0,

    "Miu3": 0,

    "F_M": f_M,

    "W_M": 0,

    "F_Miu": f_Miu,

    "W_Miu": 0,

    "X": 0,

    "Y": 0
```

```
}

# source solution

input_concentration_profile = {

"F_M": f_M,

"F_Miu": f_Miu

}

dilution_rate = 0.01

time_interval = 1

steps = 10000 # simulation steps to be run. If you want to control by
convergence rather than by simulation steps, you may modify the
simu_network_switch function to set a new stop criterion.

result = simu_network_switch(steps, initial_conc_settings, reaction_settings,
solubility_settings, dilution_rate, input_concentration_profile,
time_interval)
```